\documentclass[epj]{svjour}
\usepackage{graphicx}
\usepackage{color}

\def\mb#1{\setbox0=\hbox{$#1$}\kern-.025em\copy0\kern-\wd0
\kern-0.05em\copy0\kern-\wd0\kern-.025em\raise.0233em\box0}

\begin{document}
   \title{Statistics of the gravitational force in various dimensions of space: from Gaussian to L\'evy
    laws}

 \author{P.H. Chavanis}

\institute{Laboratoire de Physique Th\'eorique (CNRS UMR 5152),
Universit\'e Paul Sabatier, 118 route de Narbonne, \\
31062 Toulouse, France.
\email{chavanis@irsamc.ups-tlse.fr}
}

\titlerunning{Statistics of the gravitational force}

   \date{To be included later }

   \abstract{We discuss the distribution of the gravitational force
   created by a Poissonian distribution of field sources (stars,
   galaxies,...) in different dimensions of space $d$. In $d=3$, when
   the particle number $N\rightarrow +\infty$, it is given by a L\'evy
   law called the Holtsmark distribution. It presents an algebraic
   tail for large fluctuations due to the contribution of the nearest
   neighbor. In $d=2$, for large but finite values of $N$, it is given
   by a marginal Gaussian distribution intermediate between Gaussian
   and L\'evy laws. It presents a Gaussian core and an algebraic
   tail. In $d=1$, it is exactly given by the Bernouilli distribution
   (for any particle number $N$) which becomes Gaussian for $N\gg
   1$. Therefore, the dimension $d=2$ is critical regarding the
   statistics of the gravitational force. We generalize these results
   for inhomogeneous systems with arbitrary power-law density profile
   and arbitrary power-law force in a $d$-dimensional universe.
 }

   \maketitle
%

\section{Introduction}
\label{sec_introduction}

In this paper, we study the statistics of the gravitational force
arising from a random distribution of field sources (stars,
galaxies,...) in arbitrary dimensions of space $d$.  This systematic
study has interest both in astrophysics and probability theory. In
addition, the distribution of the gravitational force presents
numerous analogies with other physical systems such as Coulombian
plasmas, 2D point vortices, dislocation systems etc. Many results have
already been obtained by Chandrasekhar \cite{chandra} for the
gravitational force in $d=3$ dimensions. In view of the fundamental
nature of this problem and its potential applications in various areas
of physics and astrophysics, it is important to formulate the
mathematical problem at a general level and study how the results are
affected by the dimension of space.

If we consider stellar systems such
as globular clusters or elliptical galaxies, the problem is clearly
three-dimensional and the gravitational force between two stars scales
like $1/r^2$. The structure of self-gravitating isothermal and
polytropic spheres has been discussed by Emden \cite{emden} and
Chandrasekhar
\cite{chandrastellar} and the thermodynamics of stellar systems has
been initiated by Antonov \cite{antonov} and Lynden-Bell \& Wood
\cite{lbw}, and developed by several authors since then (see the reviews
of Padmanabhan \cite{paddy} and Chavanis \cite{ijmpb}). On the other
hand, the statistics of the gravitational force produced by a random
distribution of stars has been studied by Chandrasekhar \cite{chandra}
by analogy with the work of Holtsmark \cite{holtsmark} on the
distribution of the electrostatic field in a plasma composed of simple
ions. In a series of papers, Chandrasekhar \& von Neumann
\cite{cn0,cn1,cn2,cn3,cn4} pursued this work in order to obtain, from a
fully stochastic theory, an expression of the diffusion coefficient of
stars in a cluster and understand the origin of the logarithmic
divergence at large scales arising in the kinetic theory of stellar
systems (see Kandrup
\cite{kandrup} for a review). Numerical experiments have been
conducted by Ahmad \& Cohen \cite{ac} and more recently by Del Popolo
\cite{popolo} to test the predictions of this theory and take into account finite size effects. The initial theory was developed in the case of stars but the same methods can also be used in cosmology assuming that the field sources are galaxies rather than stars \cite{saslaw}. 

The standard results of Chandrasekhar \cite{chandra} are valid for the
gravitational force in three dimensions. However, it is important to
note that some astrophysical systems have symmetries that lead to an
effective gravitational interaction of lower dimensionality.

For example, some authors have considered the gravitational
interaction between infinitly elongated cylindrical filaments. In that
case, the force between two filaments scales like $1/r$ corresponding
to the gravitational interaction in two dimensions. The structure of
polytropic and isothermal cylinders has been studied by Ostriker
\cite{ostriker} and the thermodynamics of gravitating rods has been
developed by Katz \& Lynden-Bell \cite{klb}, Aly \& Perez \cite{ap} and
Sire \& Chavanis \cite{sc}.  Polytropic and isothermal
cylinders may have useful applications in the study of gaseous
filaments, spiral arms and rings.  Indeed, Schneider \& Elmegreen
\cite{se} have shown that dark clouds have elongated or filamentary
shapes. On the other hand, in some theoretical models, the spiral arms
of the Galaxy are considered to be self-gravitating cylinders of
infinite length \cite{cf,stodolkiewicz}. Finally, gaseous rings occur
in a variety of astronomical contexts (Saturn's ring, rings in spiral
galaxies,...)  \cite{randers,ostrikerring} and infinite cylinders
provide the first term of a natural series expansion in which one may
develop the theory of the equilibrium of such rings.

On the other hand, some authors have considered the gravitational
interaction between plane-parallel sheets. In that case, the force
between two sheets is independent on the distance, corresponding to
the gravitational force in one dimension. The isothermal and
polytropic distributions of such configurations have been determined
by Spitzer \cite{spitzer} and Camm \cite{camm} and systematically
studied by Harrison \& Lake \cite{hl} and Iba\~nez \& Sigalotti
\cite{is}. On the other hand,
 their thermodynamics has been worked out by Katz \& Lecar \cite{kl}
 and Sire \& Chavanis
\cite{sc} in the mean field approximation valid for $N\rightarrow
+\infty$. Interestingly, in the one dimensional case, Rybicki
\cite{rybicki} has shown that the statistical equilibrium state can be
calculated analytically for any $N$. Isothermal sheets can have
application in the study of galactic disks, collapsing clouds,
pancakes in cosmology and Laplacian disk cosmogony. Indeed, in
rotating disk systems, such as spiral and SO galaxies, the gas, dust
and stars tend to be distributed in a symmetrical fashion about an
equatorial plane. Camm \cite{camm} showed that the sheet model is a
useful model for stellar motion in a direction perpendicular to the
disk of a highly flattened galaxy.  On the other hand, star-forming
clouds generally collapse to a flattened (sheet), and sometimes
filamentary (cylinders), configuration before fragmenting
\cite{larson}.  Sheet-like structures may also form by interstellar
shocks or cloud collisions. Indeed, there seems to be strong evidence
that some regions of post-shocked clouds are left near
quasi-hydrostatic equilibrium plates (pancakes) at scales of galaxy
formation \cite{zeldovich} or at scales of stellar
formation in the Galaxy \cite{roberts}. Finally, plane-symmetric
distributions of matter occur in the Saturn ring system and in the
Laplacian disk cosmogony \cite{hl}.

Apart from these various physical applications, it is interesting to
investigate at a more academic level how the laws of physics, and
particularly the laws of gravity, depend on the dimension of space
$d$. There is indeed a long tradition of works in that direction
\cite{barrow} starting from a seminal paper of Ehrenfest
\cite{ehrenfest}.  For example, in Ref. \cite{wds} we have studied how the
structure of relativistic white dwarf stars would be modified in
universes with lower or higher dimensions and we found that the
dimensions $d=2$ and $d=4$ which surround the dimension $d=3$ of our
universe are {\it critical} in some respect: white dwarf stars have a
maximum radius in $d=2$, a maximum mass in $d=3$ and they become
unstable for $d\ge 4$. We have also found that the dimensions $d=2$
and $d=10$ are special for classical isothermal spheres \cite{sc} and
that the dimensions $d=2$ and $d=9.96404372...$ are special for
self-gravitating radiation in general relativity
\cite{bh}. This type of analysis can shed new light on the
anthropic principle and explain why the dimension of our universe is
particular.  This is a further motivation, in addition to the physical
examples mentioned above, to study gravity in $d$ dimensions. Extra
dimensions at the microscale also appear in theories of grand
unification and black holes, an idea originating from Kaluza-Klein
theory.

On the numerical point of view, there has been considerable interest
over the years in the behaviour of one dimensional gravitational
systems, essentially by reason of the simplicity of these models and
their relatively cheapness for numerical study.  Numerical simulations
are more easily carried out in $1D$ than in $3D$ and many early
numerical works have considered one-dimensional self-gravitating
systems (OGS) to study (i) the process of violent relaxation (ii) the
collisional evolution of the system and its relaxation to thermal
equilibrium (iii) ergodicity for gravitational systems. The OGS is
indeed the simplest model for studying $N$-body gravitational
interactions even if it is not expected to capture all the features of
$3D$ interactions. Therefore, studying gravity in one and
two dimensions can be of interest to interprete numerical
simulations. We refer to Yawn \& Miller \cite{ym} for further
references on this important topic.

The study of the distribution of the gravitational force in $d$
dimensions is also important in statistical physics and probability
theory \cite{levy,gk,feller,bg,bardou,sornette} because it is an interesting
example of a sum of random variables where the Central Limit Theorem
(CLT) may or may not apply depending on the dimension of space. In
particular, we show in this paper that the dimension $d=2$ is {\it
critical} for the statistics of the gravitational field. In $d=3$, the
variance of the gravitational force produced by one star diverges
algebraically so that the distribution of the total force is a
particular L\'evy law called the Holtsmark distribution\footnote{It
is interesting to note that Chandrasekhar (1943) \cite{chandra} did
not mention the connection between the Holtsmark distribution and
L\'evy laws. At that time, the work of L\'evy (1937) \cite{levy} was
essentially known among mathematicians and had not diffused yet in the
physical and astrophysical communities.}. It presents an algebraic
tail which is essentially due to the contribution of the nearest
neighbor. In $d=1$, the variance of the gravitational force produced
by one star is finite so that, by application of the CLT, the
distribution of the total force is Gaussian (for finite $N$ it is
exactly given by the Bernouilli distribution). In $d=2$, the variance
of the gravitational force produced by one star diverges
logarithmically so that the distribution of the total force is a
marginal Gaussian distribution intermediate between Gaussian and
L\'evy laws. It has a Gaussian core as if the CLT were applicable (but
the variance diverges logarithmically with $N$) and an algebraic tail
produced by the nearest neighbor as for a L\'evy law. Therefore, by
changing the dimension of space, we can pass from Gaussian ($d=1$) to
L\'evy ($d= 3$) laws with an interesting limit case ($d=2$). This
transition has not been reported before in the context of
gravitational dynamics and we think that it deserves a particular
discussion. 

Finally, the systematic study of the distribution of the gravitational
force in various dimensions of space is interesting in view of the
different analogies with other physical systems. In $d=3$, we have
already mentioned the analogies between the statistics of the
gravitational force created by stars in a galaxy and the statistics of
the electrostatic force created by a plasma composed of simple
ions. In $d=2$, the statistics of the gravitational force is similar
to the statistics of the velocity field created by a random
distribution of point vortices studied by Chavanis \& Sire
\cite{cs1,cs2,c3} (see also
\cite{jimenez,min,chukbar,kuv,sirec,lm}), to the statistics of the force
created by dislocations \cite{cg} or to the statistics of the
electrostatic field in a two-dimensional plasma
\cite{kind}. In $d=1$, we are not aware of any particular analogy
except with one dimensional plasmas.

This paper is organized as follows.  In Sec. \ref{sec_h}, we recall
the main lines of the calculation of the distribution of the
gravitational force in $d=3$ dimensions leading the the Holtsmark
\cite{holtsmark,chandra} distribution and discuss the main properties
of this distribution. In Sec. \ref{sec_m}, we determine the
distribution of the gravitational force in $d=2$ dimensions by
adapting the results of Chavanis \& Sire \cite{cs1} for point vortices
to the present context. In Sec. \ref{sec_g}, we determine the
distribution of the gravitational force in $d=1$ dimension. Finally,
in Sec.
\ref{sec_in}, we generalize the results of this paper to the case of
an inhomogeneous distribution of particles with arbitrary power-law
density profile (or fractal distribution) and arbitrary power-law
force in a $d$-dimensional universe. In Appendix
\ref{sec_pvg}, we give the distribution of the force created by the
nearest neighbor in arbitrary dimension $d$.  For $d\ge 2$, this
expression provides a good approximation of the tail of the true
distribution.   Concerning the notations employed in
this paper, we shall use the expression of the gravitational force in
$d$ dimensions obtained from the Poisson equation written as
$\Delta\Phi=S_d G\rho$ where $S_d$ is the surface of a unit sphere in
$d$ dimensions and $\rho$ is the density distribution. Therefore, the
force (by unit of mass) created at the origin $O$ by a single star located in ${\bf r}$ is $Gm\
{\bf r}/r^3$ in $d=3$, $Gm\ {\bf r}/r^2$ in $d=2$ and $Gm \ {\rm
sgn}(x)$ in $d=1$.  With this convention, the gravitational constant
$G$ depends on the dimension of space (it will be denoted $G_d$ in
case of ambiguity).  We shall also call the particles giving rise to
the gravitational force ``stars'' although they can be filaments,
sheets or other objects.

\section{The Holtsmark distribution in $d=3$}
\label{sec_h}

The statistics of the gravitational force in $d=3$ dimensions was
first studied by Chandrasekhar \cite{chandra} by analogy with the
statistics of the electrostatic force in a plasma studied by Holtsmark
\cite{holtsmark}. Chandrasekhar computed the probability 
density $W({\bf F})$ that a test star experiences a random force per
unity mass ${\bf F}$. He assumed that there are no correlation between
the positions of the stars (Poisson distribution) and that the medium
is infinite and homogeneous\footnote{As is well-known, an infinite and
homogeneous distribution of stars is not a steady state of a
self-gravitating system. Thus, Chandrasekhar made a sort of ``Jeans'
swindle'' \cite{bt}. However, an infinite and homogeneous distribution
of masses is justified in cosmology because the expansion of the
universe has an effect similar (in the comoving frame) to that of a
neutralizing background in plasma physics \cite{saslaw}. On the other
hand, in Chandrasekhar's study, since the distribution of the
gravitational force is dominated by the contribution of the nearest
neighbor, it is permissible to extend the size of the system to
infinity without strong influence on the distribution of the force. As
shown by Kandrup \cite{ka}, only the local density of matter close to
the star under consideration is important in determining the
distribution of the force. }. The case of a finite uniform density
distribution has been considered by Ahmad \& Cohen \cite{ac}: in that
case, the distribution of force $W_N({\bf F})$ depends on the total
number of stars $N$. On the other hand, the case of a system of
non-uniform density has been treated by Kandrup
\cite{kandrup,ka}. His study demonstrates that the basic results are
rather independent upon the density profile. Chandrasekhar \& von
Neumann have used their stochastic model to determine the speed of
fluctuations $T(F)$ \cite{cn1}, the diffusion coefficient of stars
\cite{cn1} (a calculation completed by Kandrup \cite{kandrup}) and the
spatial
\cite{cn2,cn3} and temporal \cite{cn4} correlations of the gravitational
field.

Let us consider a collection of $N$ stars with mass $m$ randomly
distributed in a sphere of radius $R$ with a uniform density
$n=3N/(4\pi R^{3})$ in average. The exact
gravitational force by unit of mass created at the center $O$ of the
domain is
\begin{equation}
{\bf F}=\sum_{i=1}^{N}{\bf f}_{i},\qquad {\bf f}_{i}={Gm \over r_{i}^{3}}{\bf r}_{i}.
\label{h1}
\end{equation}
In each realization, we choose at random the position of the stars
with a uniform distribution.  Since the positions of the individual
stars fluctuate from one realization to the other, the value of the
total force fluctuates too and we are interested by its distribution
$W({\bf F})$. The problem then consists in determining the
distribution of a sum of random variables. The distribution of the
force created by one star is obtained by writing $W({\bf f})d{\bf
f}=\tau({\bf r})d{\bf r}$ where $\tau({\bf r})=3/(4\pi R^{3})$ denotes
the density probability of finding the star in ${\bf r}$ and ${\bf
f}=Gm{\bf r}/r^3$ according to Eq. (\ref{h1}). The Jacobian of the
transformation ${\bf r}\rightarrow {\bf f}$ is readily evaluated
leading to $d{\bf f}=2(Gm)^{-3/2}f^{9/2}d{\bf r}$. Therefore, the
distribution of the individual forces is given, for $f>Gm/R^2$, by the
pure power-law
\begin{eqnarray}
W({\bf f})=\frac{1}{2}(Gm)^{3/2}\frac{3}{4\pi R^{3}}f^{-9/2}.
\label{h2}
\end{eqnarray}
The variance of the
force created by one star
\begin{eqnarray}
\langle f^{2}\rangle= {3\over 4\pi R^{3}}\int\biggl ({Gm\over r^{2}}\biggr )^{2}4\pi r^{2} dr
\propto \int_{0}^{+\infty} \frac{1}{r^2}dr
\label{h3}
\end{eqnarray}
diverges algebraically due to the behaviour at small distances
$r\rightarrow 0$ (corresponding to large forces $f\rightarrow
+\infty$). Therefore, the CLT is not applicable. As we shall see, the
distribution of the total force is a L\'evy law known as the {\it
Holtsmark distribution} since it was first determined by Holtsmark in
the context of the electric field created by a gas of simple ions
\cite{holtsmark}. We briefly summarize the procedure developed by
Chandrasekhar \cite{chandra} to compute the distribution of the
force. This summary is useful to compare with the results in other
dimensions.

Since there are no correlation between the stars, the distribution of the gravitational force for any value of $N$ can be expressed as
\begin{equation}
W_{N}({\bf  F})=\int \prod_{i=1}^{N}\tau({\bf r}_{i})d{\bf
r}_{i}\delta\left ({\bf  F}-\sum_{i=1}^{N}{\bf f}_{i}\right ),
\label{h4}
\end{equation}
where $\tau({\bf r}_{i})=3/(4\pi R^{3})$ governs the probability of
occurrence of the $i$-th star at position ${\bf r}_{i}$.  Now, using a
method originally due to Markov, we can express the $\delta$-function
appearing in Eq. (\ref{h4}) in terms of its Fourier transform. In
that case, $W_{N}({\bf F})$ becomes
\begin{equation}
W_{N}({\bf F})=\frac{1}{(2\pi)^{3}}\int A_{N}({\bf k})e^{-i {\bf k}\cdot {\bf
F}}d{\bf k},
\label{h5}
\end{equation}
with
\begin{equation}
A_{N}({\bf k})=\left (\frac{3}{4\pi R^{3}}\int_{|{\bf r}|=0}^{R}e^{i {\bf k}\cdot {\bf f}}d{\bf r}\right )^{N},
\label{h6}
\end{equation}
where we have written ${\bf f}=Gm{\bf r}/r^3$. We shall see that the distribution of the force is dominated by the contribution of the nearest neighbor. Therefore, we can consider the thermodynamic limit
\begin{equation}
N\rightarrow +\infty, \quad R\rightarrow +\infty,\quad n=\frac{3N}{4\pi R^{3}}= {\rm const.}
\label{h7}
\end{equation}
In this limit, we obtain \cite{chandra}:
\begin{equation}
W({\bf F})=\frac{1}{(2\pi)^{3}}\int A({\bf k})e^{-i {\bf k}\cdot {\bf
F}}d{\bf k},
\label{h8}
\end{equation}
with
\begin{equation}
A({\bf k})=e^{-nC({\bf k})}, \qquad C({\bf k})=\int_{|{\bf r}|=0}^{+\infty}\left (1-e^{i {\bf k}\cdot {\bf f}}\right )d{\bf r}.
\label{h9}
\end{equation}
The integral can be calculated easily \cite{chandra} leading to
\begin{equation}
A({\bf k})=e^{-a k^{3/2}}, \qquad a=\frac{4}{15}(2\pi Gm)^{3/2}n.
\label{h10}
\end{equation}
Therefore, the distribution of the gravitational force is given by the Holtsmark distribution \cite{holtsmark,chandra}:
\begin{eqnarray}
W({\bf F})={1\over 2\pi^{2}F}\int_{0}^{+\infty}e^{-ak^{3/2}}\sin (kF)k\, dk.
\label{h11}
\end{eqnarray}
It has the asymptotic behaviours \cite{chandra}:
\begin{equation}
W({\bf F})\rightarrow \frac{1}{3\pi^{2}}\left (\frac{15}{4}\right )^{2}\frac{1}{(2\pi Gm)^{3}n^2}\qquad (F\rightarrow 0),
\label{h12n}
\end{equation}
\begin{equation}
W({\bf F})\sim {1\over 2}(Gm)^{3/2}nF^{-9/2}\qquad (F\rightarrow +\infty).
\label{h12}
\end{equation}
Therefore, the variance of the gravitational force 
\begin{equation}
\langle F^{2}\rangle \propto \int^{+\infty}\frac{dF}{F^{1/2}}
\label{h13}
\end{equation}
diverges algebraically because of the contribution of large fields $F\gg 1$.
On the other hand the average value of the force is \cite{kandrup}:
\begin{eqnarray}
\langle F\rangle =4\Gamma\left (\frac{1}{3}\right)\left (\frac{8\sqrt{2}}{15}\right )^{2/3}Gm n^{2/3}\simeq 8.879 Gm n^{2/3}.\nonumber\\
\label{h14}
\end{eqnarray}
The typical force exerted upon a test particle is thus of the
magnitude $Gm/D^2$ which might be expected to arise from a few
particularly nearby field stars at the interstellar distance $D\sim
n^{-1/3}$. Writing this typical force as $F_0=Gmn^{2/3}$,
Eqs. (\ref{h12n}) and (\ref{h12}) give the asymptotic behaviour of the
Holtsmark distribution for $F\ll F_0$ and $F\gg F_0$ respectively.

\begin{figure}
\centering
\includegraphics[width=8cm]{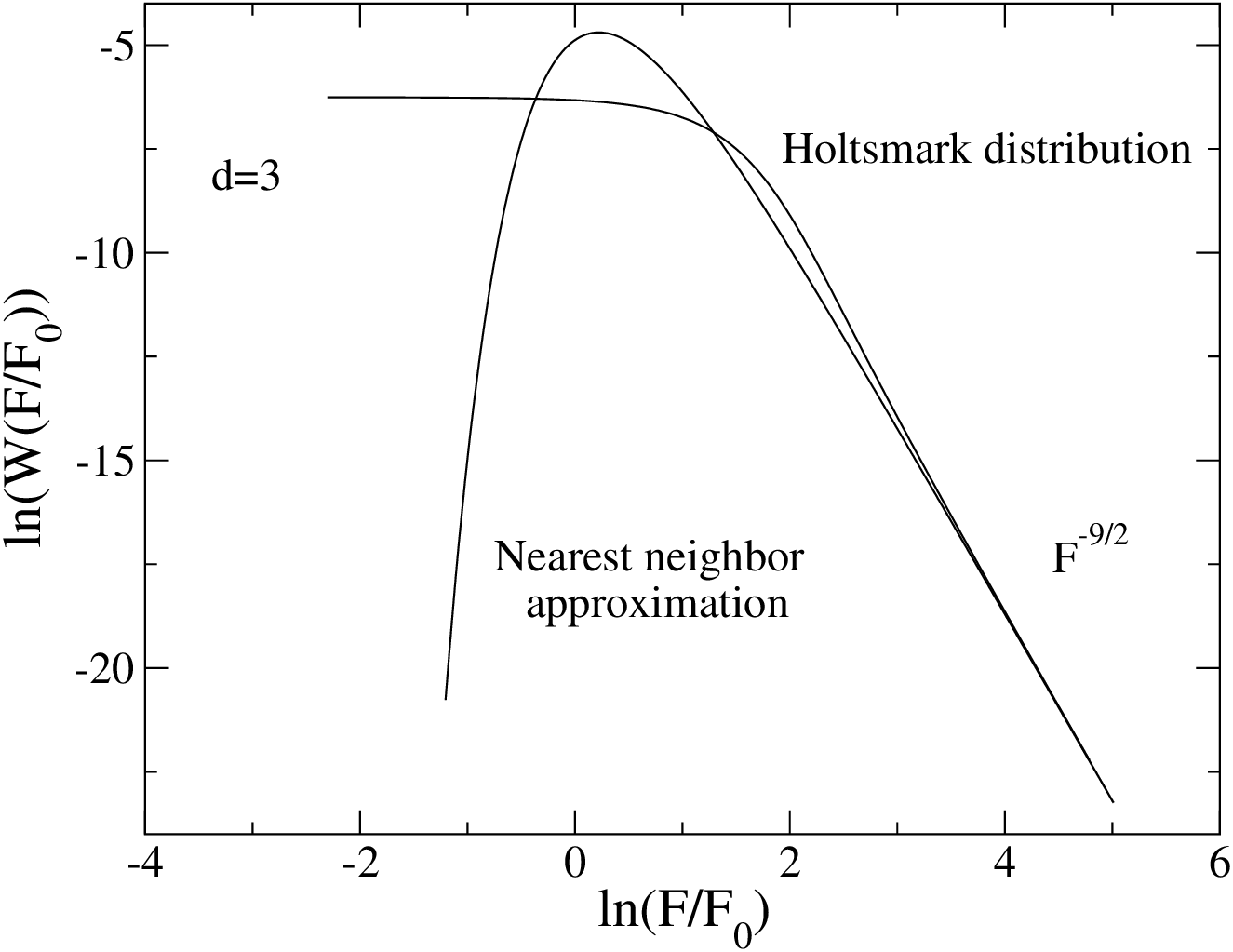}
\caption{The Holtsmark distribution in $d=3$. It is compared with the distribution of the force due to the nearest neighbor. This clearly shows that the tail of the distribution is produced by the nearest neighbor. We have set $F_{0}=Gmn^{2/3}$.}
\label{holtsmark}
\end{figure}

It is instructive to compare the Holtsmark distribution (\ref{h11})
with the distribution of the force created by the nearest neighbor
given by \cite{chandra}:
\begin{equation}
W_{n.n.}({\bf F})=\frac{1}{2}(Gm)^{3/2}nF^{-9/2}e^{-\frac{4\pi (Gm)^{3/2}n}{3F^{3/2}}}.
\label{h15}
\end{equation}
It has the asymptotic behaviour
\begin{equation}
W_{n.n.}({\bf F})\sim {1\over 2}G^{3/2}m^{3/2}nF^{-9/2}\qquad (F\rightarrow +\infty),
\label{h16}
\end{equation}
which is in {exact} agreement with the asymptotic behaviour
(\ref{h12}) of the Holtsmark distribution. Therefore, the highest
fields are produced only by the nearest neighbor. By contrast, in the
limit of weak forces, the two distributions disagree: whereas the
nearest neighbor distribution vanishes exponentially, the Holtsmark
distribution tends to a constant value (\ref{h12n}). This reflects the
fact that in the case of extremely weak forces, more than one field
star plays a significant role. The Holtsmark distribution is compared
to the distribution of the force due to the nearest neighbor in
Fig. \ref{holtsmark} and we get a good agreement for sufficiently
large forces $F\gg F_0$.  The typical force due to the nearest
neighbor is $F_{n.n}\sim Gm/D^{2}\sim Gmn^{2/3}$ where $D$ is the
average distance between stars. It is precisely of the same order as
the average value of the force (\ref{h14}) due to all the stars.  More
precisely, the average value of the force due to the nearest neighbor
is \cite{kandrup}:
\begin{eqnarray}
\langle F\rangle_{n.n.}= \Gamma\left (\frac{1}{3}\right)\left (\frac{4\pi}{3}\right )^{2/3}Gm n^{2/3}\simeq 6.926 Gm n^{2/3}.\nonumber\\
\label{h17}
\end{eqnarray}
The two results (\ref{h14}) and (\ref{h17}) differ only by a factor
$1.28$. Therefore, large forces are due essentially to the
contribution of the nearest neighbor and the effect of all other field
stars cancels. Thus, field stars separated by distances large compared
with the interstellar spacing contribute negligibly to the total
stochastic force. This is the reason why Chandrasekhar \& von Neumann
\cite{cn1} argue that the logarithmic divergence in the diffusion
coefficient of stars has to be cut-off at the inter-spacing distance
$D$\footnote{This argument is controversial because it would imply that
the logarithmic divergence in the diffusion coefficient of charges in
a plasma must also be cut-off at the inter-spacing distance while the
works of Cohen {\it et
al.}
\cite{cohen} and Balescu \cite{balescu} show that it has to be cut-off at the
(larger) Debye length. Therefore, for a stellar system, the
logarithmic divergence in the diffusion coefficient of stars should be
cut-off at the system's size $R$, or Jeans length, which is the
gravitational analogue of the Debye length in plasma physics.}.

Chandrasekhar's approach shows that only stars close to the star under
consideration determine the fluctuations of the gravitational
field (for large forces). In fact, Agekyan \cite{agekyan} has calculated the average
contribution to the total magnitude of the stochastic gravitational
field due to some field star at a distance $r$ from the star under
consideration. He finds that it produces  an effective force which in good approximation 
can be written as \cite{agekyan}:
\begin{equation}
f_{eff}={Gm\over r^{2}}{1\over 1+{r^{2}/\Lambda^{2}}},
\label{h18}
\end{equation}
where 
\begin{equation}
\Lambda=\left\lbrack \frac{4\Gamma(\frac{2}{3})}{9\pi}\left (\frac{3}{4\pi}\right )^{2/3}\right\rbrack^{1/2} n^{-1/3}=0.271 \ n^{-1/3},
\label{h18n}
\end{equation}
is a characteristic length of the order of the interparticle distance
$D$.  For weak separations, one has $f_{eff}\rightarrow Gm/r^{2}$ but
for large separations $r\gg D$, the effects of individual stars
compensate each other and the resulting force is reduced by a factor
$(r/D)^{2}$.  This corroborates the heuristic argument of
Chandrasekhar \cite{chandra} concerning the validity of the two-body
approximation and the effective cancellation of the contribution from
distant field stars. If we were naively to assume that the average
value of the {\it modulus} of the force is additive $\langle |{\bf
F}|\rangle=\sum_{i=1}^{N}\langle |{\bf f}_i|\rangle$ we would find the
wrong result that the average force is infinite: $\langle
F\rangle=N\langle f\rangle=n\int_0^{+\infty}\frac{Gm}{r^2}4\pi r^2
dr=+\infty$. In fact, we must sum the modulus of the {\it effective}
force $\langle |{\bf F}|\rangle=\sum_{i=1}^{N}\langle |({\bf
f}_{eff})_i|\rangle$ and since the effective force decreases like
$1/r^4$, we find a finite value. If we write
\begin{equation}
\langle F\rangle=N\langle f_{eff}\rangle=n\int_0^{+\infty} \frac{Gm}{r^2}\frac{1}{1+r^2/\Lambda^2} 4\pi r^2 dr
\label{h18v}
\end{equation}
we
obtain 
\begin{eqnarray}
\langle F\rangle=2\pi^2\left\lbrack \frac{4\Gamma(\frac{2}{3})}{9\pi}\left (\frac{3}{4\pi}\right )^{2/3}\right\rbrack^{1/2} Gmn^{2/3}=5.349 G m n^{2/3},\nonumber\\
\label{h18nn}
\end{eqnarray}
which is close to the exact result (\ref{h14}). On the other hand, the
average value of the squared force is clearly additive since the
particles are uncorrelated
\begin{equation}
\langle F^2\rangle=\langle \sum_{ij} {\bf f}_i \cdot {\bf f}_j\rangle= \sum_{i=1}^{N} \langle f_i^2\rangle+ \sum_{i\neq j} \langle {\bf f}_i\rangle \cdot \langle {\bf f}_j\rangle=\sum_{i=1}^{N} \langle f_i^2\rangle. 
\label{h18nnn}
\end{equation}
Therefore, we find that the variance of the gravitational force is infinite 
\begin{equation}
\langle F^2\rangle=N\langle f^2\rangle=n\int_0^{+\infty} \left (\frac{Gm}{r^2}\right )^2 4\pi r^2 dr=+\infty
\label{h18nnnn}
\end{equation}
in agreement with Eq. (\ref{h13}).

\section{The marginal Gaussian distribution in $d=2$}
\label{sec_m}

The gravitational field produced in $O$ by an infinite rod of mass per
unit length $\mu$ is ${\bf f}=2G\mu {\bf r}/r^2$. This corresponds to
the gravitational force ${\bf f}=G_{2}m {\bf r}/r^2$ created by a mass
$m$ in two dimensions provided that we make the correspondance
$G_{2}=2G\mu/m$. Now, the statistics of the gravitational force in
$d=2$ dimensions can be directly obtained from the work of Chavanis \&
Sire \cite{cs1} on the statistics of the velocity created by a random
distribution of point vortices in 2D hydrodynamics (it suffices to
make the correspondance $\gamma/(2\pi)\leftrightarrow Gm$ where
$\gamma$ is the circulation of a point vortex). There are indeed
remarkable analogies between stellar systems and 2D vortices
\cite{houches}. Adapting the procedure of Chandrasekhar
\& von Neumann \cite{cn1}, Chavanis \& Sire \cite{cs1,cs2,c3} have
used this stochastic approach to obtain an estimate of the diffusion
coefficient of point vortices when their distribution is homogeneous.
This is another manifestation of the deep formal analogy between stars
and galaxies. Other related works on the statistics of the velocity
created by point vortices, including direct numerical simulations to
test the theoretical results, have been performed in
\cite{jimenez,min,chukbar,kuv,sirec,lm}.

Let us consider a collection of $N$ particles with mass $m$ randomly
distributed in a disk of radius $R$ with a uniform density
$n=N/(\pi R^{2})$ in average. The force by unit of mass created at
the center $O$ of the domain is
\begin{equation}
{\bf F}=\sum_{i=1}^{N}{\bf f}_{i},\qquad {\bf f}_{i}={Gm \over r_{i}^{2}}{\bf r}_{i}.
\label{m1}
\end{equation}
The problem consists in determining the distribution of a
sum of random variables. The distribution of the force created by one star is
obtained by writing $W({\bf f})d{\bf f}=\tau({\bf r})d{\bf r}$ where
$\tau({\bf r})=1/(\pi R^{2})$ denotes the density probability of
finding the star in ${\bf r}$. Using $d{\bf f}=(Gm)^{-2}f^4 d{\bf r}$, we obtain, for $f>Gm/R$, the pure power-law:
\begin{eqnarray}
W({\bf f})=(Gm)^{2}\frac{1}{\pi R^{2}}f^{-4}.
\label{m2}
\end{eqnarray}
The variance of the
force created by one star
\begin{eqnarray}
\langle f^{2}\rangle= {1\over \pi R^{2}}\int\biggl ({Gm\over r}\biggr )^{2}2\pi r dr
\propto \int_{0}^{+\infty} \frac{dr}{r}
\label{m3}
\end{eqnarray}
diverges logarithmically due to the behaviour at small and large
distances (corresponding to weak $f\rightarrow 0$ and large
$f\rightarrow +\infty$ forces). Therefore, strictly speaking, the CLT
is not applicable. However, since the divergence of the variance is
weak (logarithmic) we shall see that the distribution of the total
force is intermediate between Gaussian and L\'evy laws. The core of
the distribution is Gaussian as if the CLT were applicable (but the
variance diverges logarithmically with $N$) while the tail is
algebraic, and produced by the nearest neighbor, as for a L\'evy law.

Following the method previously exposed and considering the
thermodynamic limit
\begin{equation}
N\rightarrow +\infty, \quad R\rightarrow +\infty,\quad n=\frac{N}{\pi R^{2}}= {\rm const.}
\label{m4}
\end{equation}
we obtain
\begin{equation}
W({\bf F})=\frac{1}{(2\pi)^{2}}\int A({\bf k})e^{-i {\bf k}\cdot {\bf
F}}d{\bf k},
\label{m5}
\end{equation}
with
\begin{equation}
A({\bf k})=e^{-nC({\bf k})}, \qquad C({\bf k})=\int_{|{\bf r}|=0}^{R}\left (1-e^{i {\bf k}\cdot {\bf f}}\right )d{\bf r},
\label{m6}
\end{equation}
where we have written ${\bf f}=Gm{\bf r}/r^2$. Note that we cannot let
$R\rightarrow +\infty$ in the last integral since it diverges
logarithmically for large $r$. Still, the procedure is well-defined
mathematically if we view (\ref{m5})-(\ref{m6}) as an {\it equivalent}
of $W_{N}({\bf F})$ for large $N$, not a true limit. The integral in (\ref{m6}) can be calculated explicitly \cite{cs1} leading to
\begin{equation}
A({\bf k})=e^{-a k^{2}\ln \left (\frac{N}{4ak^{2}}\right )}, \qquad a=\frac{1}{4}\pi (Gm)^{2}n.
\label{m7}
\end{equation}
For $F<F_{crit}(N)$ where $F_{crit}(N)$ is defined by Eq. (\ref{m12}),
we need to consider large values of $k$ in Eq. (\ref{m7}) and we can
neglect the contribution of $k$ in the logarithm, writing $A({\bf
k})\simeq e^{-a k^{2}\ln N}$. Therefore, we get a Gaussian
distribution
\begin{eqnarray}
W({\bf F})=\frac{1}{n(Gm)^2\pi^{2}\ln N}e^{-\frac{F^{2}}{n(Gm)^2\pi\ln N}}\quad (F<F_{crit}(N)),\nonumber\\
\label{m8}
\end{eqnarray}
as if the CLT were applicable. However, if we were to extend this distribution for all values of $F$, we see that the variance of this distribution
\begin{eqnarray}
\langle F^{2}\rangle = n(Gm)^{2}\pi\ln N
\label{m9}
\end{eqnarray}
diverges logarithmically with $N$ due to cooperative effects. On the other hand, the average value of the force is
\begin{equation}
\langle F\rangle =\left (\frac{1}{4}nG^2m^2\pi^2\ln N\right )^{1/2}.
\label{m13}
\end{equation}
For $F>F_{crit}(N)$, we need to consider small values of $k$ in
Eq. (\ref{m7}) and its contribution in the logarithm becomes crucial,
so that $A({\bf k})\simeq e^{2a k^{2}\ln k}$. In that case, we find
after some calculation \cite{cs1} that
\begin{eqnarray}
W({\bf F})=n(Gm)^{2}F^{-4}\qquad (F>F_{crit}(N)).
\label{m10}
\end{eqnarray}
Therefore, the distribution of the gravitational field in $d=2$ has an
algebraic tail as for a L\'evy law. The variance
of the gravitational force
\begin{equation}
\langle F^{2}\rangle \propto \int^{+\infty}\frac{dF}{F}
\label{m11}
\end{equation}
diverges logarithmically due to the contribution of large field
strengths. Comparing Eqs.  (\ref{m8}) and (\ref{m10}), we find that
the typical force where the two regimes (Gaussian core and algebraic
tail) connect each other is
\begin{eqnarray}
F_{crit}(N)\sim (nG^{2}m^{2}\pi\ln N)^{1/2}\ln^{1/2}(\ln N).
\label{m12}
\end{eqnarray}
For $N\rightarrow +\infty$, $F_{crit}(N)\rightarrow +\infty$ so,
strictly speaking, the algebraic tail is rejected to infinity and
the limit distribution $W({\bf F})$ is Gaussian. However, for large but
finite values of $N$, the convergence to the limit distribution is so
slow that the algebraic tail is always visible in practice. The
contribution of the high field tail to the average value of the force
is
\begin{eqnarray}
\langle F\rangle=2\pi n (Gm)^2\int_{F_{crit}
}^{+\infty}\frac{d F}{F^2}=\left (\frac{4\pi n G^2 m^2}{\ln N \ln (\ln N)}\right )^{1/2},\nonumber\\
\label{m12n}
\end{eqnarray}
which is smaller than the contribution (\ref{m13}) due to the core of
the distribution.

\begin{figure}
\centering
\includegraphics[width=8cm]{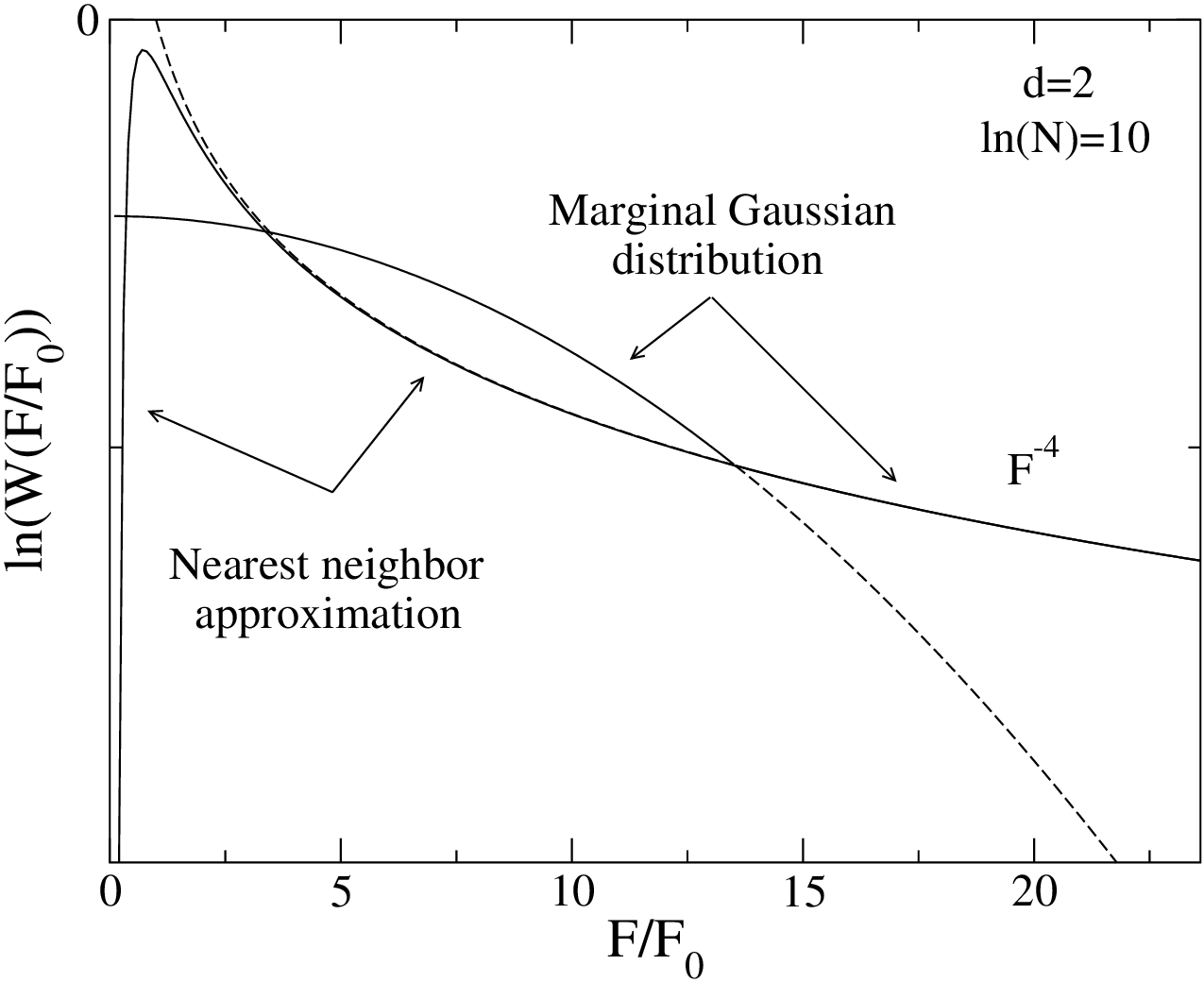}
\caption{The marginal Gaussian distribution in $d=2$. It is compared with the distribution of the force due to the nearest neighbor. This illustrates the fact that the tail of the distribution is produced by the nearest neighbor. We have set $F_{0}=Gmn^{1/2}$.}
\label{marginal}
\end{figure}

It is instructive to compare the marginal Gaussian  distribution with the distribution of the force created by the nearest neighbor given by \cite{cs1}:
\begin{equation}
W_{n.n.}({\bf F})=n(Gm)^{2}F^{-4}e^{-\frac{\pi (Gm)^{2}n}{F^{2}}}.
\label{m14}
\end{equation}
It has the asymptotic behaviour
\begin{equation}
W_{n.n.}({\bf F})\sim n(Gm)^{2}F^{-4}\qquad (F\rightarrow +\infty),
\label{m15}
\end{equation}
which is in exact agreement with the asymptotic behaviour (\ref{m10})
of the marginal Gaussian distribution.  Therefore, the highest fields
are produced only by the nearest neighbor as for a L\'evy
law. However, the Gaussian distribution of the core is created by all
the particles so that the distribution (\ref{m14}) does not provide a
good approximation of the distribution for intermediate values of the
force. The marginal Gaussian distribution is compared to the
distribution of the force due to the nearest neighbor in
Fig. \ref{marginal} and we get a good agreement only in the tail of
the distribution.  The typical force due to the nearest neighbor is
$F_{n.n}\sim Gm/D\sim Gmn^{1/2}$ where $D\sim n^{-1/2}$ is the average
distance between stars.  More precisely, the average value of the
force due to the nearest neighbor is
\begin{equation}
\langle F\rangle_{n.n.} =\pi G m n^{1/2}.
\label{m16}
\end{equation}
It is less than the average value of the force (\ref{m13}) due to all
the stars because of the $\ln N$ factor arising from cooperative
effects. However, apart from this logarithmic term, they are of the
same order of magnitude. This means that the force created by the
nearest neighbor is of the same order as the force due to all the
other particles (up to a logarithmic correction). This is another
manifestation of the fact that we lie at the frontier between Gaussian
and L\'evy laws.

The present  approach shows that only stars close to the star under
consideration determine the fluctuations of the gravitational field
(for large forces). In fact, by adapting the calculations of Agekyan
in $d=2$, it is possible to show that the ``effective'' force created
by a star at distance $r$ from the star under consideration is given
in good approximation by \cite{c3}:
\begin{equation}
f_{eff}={Gm\over r}{1\over 1+{r/\Lambda}},
\label{m17}
\end{equation}
where 
\begin{equation}
\Lambda=(16n\ln N)^{-1/2}
\label{m17n}
\end{equation}
is of the order of the interparticle distance $D$.
For weak separations, one has $f_{eff}\rightarrow Gm/r$ but for
large separations $r\gg D$, the effects of individual stars compensate
each other and the resulting force is reduced by a factor
$(r/D)$. The average value of the modulus of the force can be obtained by summing
the modulus of the effective force writing 
\begin{equation}
\langle F\rangle=N\langle f_{eff}\rangle=n\int_0^{+\infty} \frac{Gm}{r}\frac{1}{1+r/\Lambda} 2\pi r dr.
\label{h18vb}
\end{equation}
This yields
\begin{equation}
\langle F\rangle =\left (\frac{1}{16}nG^2m^2\pi^2\ln N\right )^{1/2},
\label{m13b}
\end{equation}
which is comparable to the exact result (\ref{m13}). On the other hand, the variance of the gravitational force is infinite
\begin{equation}
\langle F^2\rangle=N\langle f^2\rangle=n\int_0^{+\infty} \left (\frac{Gm}{r}\right )^2 2\pi r dr=+\infty
\label{h18nnnnb}
\end{equation}
in agreement with Eq. (\ref{m11}).

\section{The Gaussian distribution in $d=1$}
\label{sec_g}

The gravitational field produced in $O$ by an infinite sheet of mass
per unit surface $\mu$ is ${f}=2\pi G\mu \ {\rm sgn}(x)$. This
corresponds to the gravitational force ${f}=G_{1}m \ {\rm sgn}(x)$
created by a mass $m$ in one dimension provided that we make the
correspondance $G_{1}=2\pi G\mu/m$. As we have mentioned in the
Introduction, the one dimensional self-gravitating system (OGS) has
been suggested as a model for the motion of stars perpendicular to the
plane of highly flattened disk galaxies and it has been extensively
studied in numerical simulations for its computational ease \cite{ym}.

The statistics of the gravitational force in $d=1$ dimension created
by a Poissonian distribution of stars is relatively
straightforward. Let us consider a collection of $N$ particles with
mass $m$ randomly distributed in an interval $\lbrack -L,+L\rbrack$
with a uniform density $n=N/(2L)$ in average. The force by unit of
mass created at the center $O$ of the domain is
\begin{equation}
{F}=\sum_{i=1}^{N}{f}_{i},\qquad {f}_{i}=Gm \ {\rm sgn}(x_{i}),
\label{g1}
\end{equation}
where ${\rm sgn}(x)=+1$ if $x>0$ and ${\rm sgn}(x)=-1$ if $x<0$. The
total force in $O$ can be written
\begin{equation}
{F}=Gm(N_{+}-N_{-}),
\label{g2}
\end{equation}
where $N_{+}$ is the number of stars in the interval $0<x\le L$ and  $N_{-}$ is the number of stars in the interval $-L\le x<0$. Since $N_{+}+N_{-}=N$, we can rewrite Eq. (\ref{g2}) as
\begin{equation}
{F}=Gm(2N_{+}-N).
\label{g3}
\end{equation}
We note that, in one dimension, the gravitational force takes only
{\it discrete} values. On the other hand, according to Eq. (\ref{g3}),
the probability of the fluctuation $F$ is equal to the probability of
having $N_{+}$ stars in the interval $0<x\le L$. Since the $N$
particles are uniformly distributed in the domain of size $2L$, the
probability that a star is in the interval $0<x\le L$ is
$p=L/(2L)=1/2$ and the probability that a star is in the interval
$-L\le x<0$ is $q=1-p=L/(2L)=1/2$. Therefore, the probability to have
$N_{+}$ stars in the interval $0<x\le L$ is given by the
Bernouilli distribution:
\begin{equation}
W_{N}(N_{+})=\frac{N!}{N_{+}!(N-N_{+})!}\left (\frac{1}{2}\right)^{N}.
\label{g4}
\end{equation}
The first two moments of this distribution are $\langle
N_{+}\rangle=N/2$ and $\langle (N_{+}-\langle
N_{+}\rangle)^2\rangle=N/4$. Using Eq. (\ref{g3}), the
distribution of the gravitational force in $d=1$ is exactly (i.e. for
any $N$) given by the Bernouilli distribution
\begin{equation}
W_{N}(F)=\frac{N!}{\left (\frac{N}{2}+\frac{F}{2Gm}\right )!\left(\frac{N}{2}-\frac{F}{2Gm}\right )!}\frac{1}{2^N},
\label{g5}
\end{equation}
with
\begin{equation}
\langle F\rangle=0, \qquad \langle F^2\rangle =NG^2m^2.
\label{g6}
\end{equation}
This last result  can be obtained without computation
since the variance of the force created by one particle is finite and
given by
\begin{equation}
\langle f^2\rangle=G^2m^2.
\label{g7}
\end{equation}
Since the particles are uncorrelated, and since $\langle f\rangle =0$, we have $\langle F^2\rangle=\sum_{ij}\langle f_i f_j\rangle=\sum_{i=1}^{N}\langle f_i^2\rangle+\sum_{i\neq j}\langle f_{i}\rangle \langle f_{j}\rangle=N\langle f^2\rangle=NG^2m^2$.

\begin{figure}
\centering
\includegraphics[width=8cm]{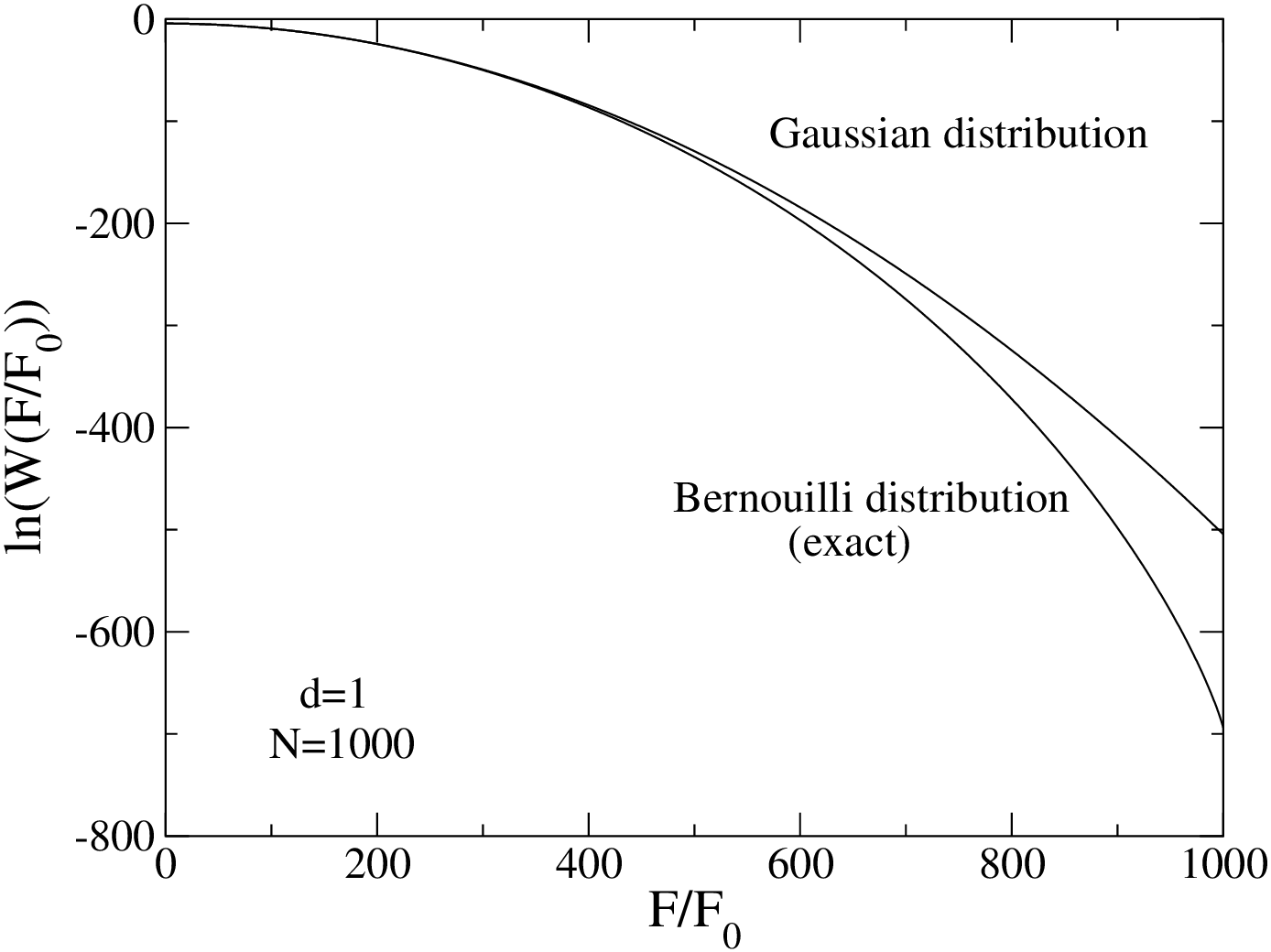}
\caption{The Gaussian distribution $W_{\infty}(F)=2GmW(F)$ in $d=1$. It is compared with the exact Bernouilli distribution $W_{N}(F)$ valid for any $N$.  We have set $F_{0}=Gm$ and taken $N=1000$. }
\label{bernouilli}
\end{figure}

For $N\gg 1$, the CLT applies and we get the Gaussian distribution
\begin{equation}
W(F)=\frac{1}{\sqrt{2\pi NG^2m^2}}e^{-F^2/(2NG^2m^2)}.
\label{g8}
\end{equation}
In the limit $N\rightarrow +\infty$, the natural scaled variable is
$F/\sqrt{N}$. The thermodynamic limit $N\rightarrow +\infty$,
$L\rightarrow +\infty$ with $N/L$ fixed is not valid in $d=1$. The
result (\ref{g8}) can also be obtained from the Bernouilli
distribution (\ref{g4}) which becomes Gaussian in the limit of large
numbers
\begin{equation}
W(N_+)\simeq \left (\frac{2}{N\pi}\right )^{1/2}e^{-\frac{2}{N}(N_+-N/2)^2}.
\label{g9}
\end{equation}
The comparison between the Bernouilli distribution and the Gaussian
distribution is shown in Fig. \ref{bernouilli}. Note that in
Eqs. (\ref{g8}) and (\ref{g9}) the particle number $N_+$ and the
gravitational force $F$ are treated as continuous variables so that
the normalization conditions are $\int_{-\infty}^{+\infty} W(F)dF=1$
and $\int_{-\infty}^{+\infty} W(N_{+})dN_{+}=1$ while in
Eqs. (\ref{g4}) and (\ref{g5}) the particle number $N_+$ and the
gravitational force $F/Gm$ are discrete variables so that the
normalization conditions are $\sum_{\frac{F}{Gm}=-N}^{N} W_{N}(F)=1$
and $\sum_{N_{+}=0}^{N} W_{N}(N_{+})=1$. Therefore, the relations
between the discrete and the continuous distributions are
$W_{N}(N_+)=W(N_+)dN_+$ with $dN_+\simeq 1$ (so that
$W_{N}(N_+)=W(N_+)$) and $W_{N}(F)=W(F)dF$ with $dF=2Gm dN_{+}\simeq
2Gm$ (so that $W_{N}(F)=2GmW(F)$).

In the preceding calculations, we have assumed that the distribution
of stars is spatially homogeneous and we have focused on the force at
the center of the domain. Since the force only depends on the number
of stars in the left and right intervals, and not on their precise
distribution, the above results remain valid for any symmetrical
distribution of the stars. If the distribution is not symmetric with
respect to the point under consideration, we just have to compute the
probability $p$ of finding a star in the right interval and use the
general Bernouilli formula. To be specific, consider an arbitrary
distribution of stars with numerical density $n(x')$ in the interval
$\lbrack L_{min},L_{max}\rbrack$.  We are interested in the
distribution of the gravitational force at $x$. The probability that a
star is in the right interval $\lbrack x,L_{max}\rbrack$ is
\begin{equation}
p(x)=\frac{1}{N}\int_{x}^{L_{max}}n(x')\, dx',
\label{g10}
\end{equation} 
and its probability to be in the left interval $\lbrack L_{min},x
\rbrack$ is $q(x)=1-p(x)$. For example, if the stars are uniformly distributed in
the interval $\lbrack -L,L\rbrack$, we have
$p(x)=\frac{1}{2}(1-\frac{x}{L})$ and
$q(x)=\frac{1}{2}(1+\frac{x}{L})$. Now, the probability to
have $N_{+}$ stars in the right interval is given by the Bernouilli
distribution:
\begin{equation}
W_{N}(N_{+},x)=\frac{N!}{N_{+}!(N-N_{+})!}p(x)^{N_{+}}q(x)^{N-N_{+}}.
\label{g11}
\end{equation}
The first two moments of this distribution are $\langle
N_{+}\rangle=Np$ and $\langle (N_{+}-\langle
N_{+}\rangle)^2\rangle=Npq$. Using Eq. (\ref{g3}), the
exact distribution of the gravitational force in $x$ is 
\begin{equation}
W_{N}(F,x)=N!\frac{p(x)^{\frac{N}{2}+\frac{F}{2Gm}}q(x)^{\frac{N}{2}-\frac{F}{2Gm}}}{\left (\frac{N}{2}+\frac{F}{2Gm}\right )!\left(\frac{N}{2}-\frac{F}{2Gm}\right )!},
\label{g12}
\end{equation}
with
\begin{equation}
\langle F\rangle=NGm(2p-1), 
\label{g13}
\end{equation}
\begin{equation}
\langle (F-\langle F\rangle)^{2}\rangle =4NG^{2}m^2pq.
\label{g14}
\end{equation}
The results (\ref{g13})-(\ref{g14}) can be obtained without
computation. The distribution of the gravitational force created created by one star is $W(f)=0$ if $f\neq \pm Gm$,  $W(f)=p(x)$ if $f=Gm$ and $W(f)=1-p(x)=q(x)$ if $f=-Gm$. The average value and the variance of the force created
by one star are given by
\begin{equation}
\langle f\rangle =pGm+(1-p)(-Gm)=(2p-1)Gm,
\label{g15}
\end{equation}
\begin{equation}
\langle f^2\rangle=p(Gm)^2+(1-p)(-Gm)^{2}=G^{2}m^{2}.
\label{g16}
\end{equation}
Now, $\langle F\rangle=\sum_{i=1}^{N}\langle f_i \rangle=N\langle
f\rangle$ and, since the particles are uncorrelated, $\langle
F^2\rangle=\sum_{ij}\langle f_i f_j\rangle=\sum_{i=1}^{N}\langle
f_i^2\rangle+\sum_{i\neq j}\langle f_{i}\rangle \langle
f_{j}\rangle=N\langle f^2\rangle+N(N-1)\langle
f\rangle^{2}$. Combining the previous results, we immediately obtain
Eqs. (\ref{g13}) and (\ref{g14}). Finally, for $N\gg 1$, the CLT
applies and we get the Gaussian distribution
\begin{equation}
W(F)=\frac{1}{\sqrt{2\pi \langle (F-\langle F\rangle)^{2}\rangle}}e^{-\frac{(F-\langle F\rangle)^2}{2\langle (F-\langle F\rangle)^{2}\rangle}}.
\label{g17}
\end{equation}
The result (\ref{g17}) can also be
obtained from the Bernouilli distribution (\ref{g12}) which becomes
Gaussian in the limit of large numbers.

\section{Inhomogeneous medium and power-law potential in $d$ dimensions}
\label{sec_in}

In this section, we determine the distribution of the force created
by an inhomogeneous distribution of particles in $d$ dimensions. To be
specific, we consider a power-law density profile $n({\bf r})=K/r^p$ for
$r\le R$ and $n({\bf r})=0$ for $r>R$. We assume $0\le p<d$ in order to have
a decreasing density distribution that is normalizable as
$r\rightarrow 0$. Then, $K=(d-p)N/(S_{d}R^{d-p})$. The uniform profile
is recovered for $p=0$ and $K=n$. For the sake of generality, we consider
a force of the form
\begin{equation}
{\bf f}=Gm\frac{{\bf r}}{r^{(d+\alpha)}},
\label{in1}
\end{equation}
with $d+\alpha-1>0$. The gravitational force is recovered for
$\alpha=0$. The distribution of the force created by one particle is obtained by writing $W({\bf f})d{\bf f}=\tau({\bf r})d{\bf r}$ where $\tau({\bf r})=(K/N)r^{-p}$. Using the transformation
\begin{equation}
d{\bf f}=(d+\alpha-1)(Gm)^{-\frac{d}{d+\alpha-1}}f^{\frac{d(d+\alpha)}{d+\alpha-1}}\, d{\bf r},
\label{df3}
\end{equation}
we obtain, for $f>Gm/R^{d+\alpha-1}$, a pure power law
\begin{equation}
W({\bf f})=\frac{1}{d+\alpha-1}\frac{d-p}{S_{d}R^{d-p}}(Gm)^{\frac{d-p}{d+\alpha-1}}f^{-\frac{d(d+\alpha)-p}{d+\alpha-1}},
\label{mon1}
\end{equation}
decreasing with an exponent
\begin{equation}
\gamma=d+\frac{d-p}{d+\alpha-1}.
\label{mon2}
\end{equation}
This distribution is normalizable provided that $\gamma>d$ which is
equivalent to our previous assumptions.  For the gravitational force
($\alpha=0$) and for a homogeneous distribution of stars ($p=0$), one
has $\gamma=d^2/(d-1)$ for $d>1$.  We note that the variance of the force
created by one star
\begin{equation}
\langle f^2\rangle\propto \int_{0}^{+\infty} \frac{1}{r^{2(d+\alpha-1)}}\times \frac{1}{r^p}\times r^{d-1}dr\propto  \int_{0}^{+\infty} \frac{dr}{r^{d+2\alpha-1+p}}
\label{in2}
\end{equation}
diverges algebraically for
\begin{equation}
2\alpha+p> 2-d,
\label{in3}
\end{equation}
due to the behaviour at small distances $r\rightarrow 0$. In that
case, the distribution of the (total) force is a L\'evy law.  For a
uniform distribution ($p=0$), the criterion (\ref{in3}) gives $\alpha>
(2-d)/2$ and for the gravitational case ($\alpha=0$) it gives
$p>2-d$. For a uniform distribution and a gravitational force
($p=\alpha=0$), we recover the condition $d>2$. In the following, we
assume that inequality (\ref{in3}) is fulfilled. The critical case
where (\ref{in3}) is an equality, corresponding to a logarithmic
divergence of the variance, will be treated specifically in
Sec. \ref{sec_margw}.

These criteria can also be expressed in terms of the index $\gamma$ of the individual distribution (\ref{mon1}). For a given dimension of space $d$, we introduce the critical exponent $\gamma_{c}=2+d$.
The variance of $W({\bf f})$ diverges algebraically (due to its
behaviour for large $f$) when $\gamma<\gamma_{c}$. In that case,
the distribution $W({\bf F})$ is a L\'evy law. For the critical case
$\gamma=\gamma_{c}$, the variance of $W({\bf f})$ diverges
logarithmically and $W({\bf F})$ is a marginal Gaussian
distribution. For $\gamma>\gamma_{c}$, the variance of $W({\bf f})$
is finite and $W({\bf F})$ is a Gaussian distribution. In the following, we assume
\begin{equation}
d<\gamma<\gamma_{c}=2+d
\label{mon3}
\end{equation} 
and in Sec. \ref{sec_margw} we consider the critical case
$\gamma=\gamma_{c}$.  

\subsection{The distribution of the force}
\label{sec_df}

We wish to determine the distribution of the total force
\begin{equation}
{\bf F}=\sum_{i=1}^{N}{\bf f}_{i},
\label{mon4}
\end{equation}
created by the particles. Since there are no correlation between the particles, the distribution of the gravitational force for any value of $N$ is given by
\begin{equation}
W_{N}({\bf  F})=\int \prod_{i=1}^{N}\tau({\bf r}_{i})d{\bf
r}_{i}\delta\left ({\bf  F}-\sum_{i=1}^{N}{\bf f}_{i}\right ),
\label{lo1}
\end{equation}
where $\tau({\bf r}_{i})$ governs the probability of
occurrence of the $i$-th star at position ${\bf r}_{i}$.  Now, using the Markov method, we express the $\delta$-function
appearing in Eq. (\ref{lo1}) in terms of its Fourier transform
\begin{equation}
\delta({\bf x})=\frac{1}{(2\pi)^{d}}\int e^{-i{\bf k}\cdot {\bf x}}\, d{\bf k}.
\label{lo2}
\end{equation}
Then, $W_{N}({\bf F})$ can be written
\begin{equation}
W_{N}({\bf F})=\frac{1}{(2\pi)^{d}}\int A_{N}({\bf k})e^{-i {\bf k}\cdot {\bf
F}}d{\bf k},
\label{lo3}
\end{equation}
with
\begin{equation}
A_{N}({\bf k})=\left (\int_{|{\bf r}|=0}^{R}\tau({\bf r}) e^{i {\bf k}\cdot {\bf f}}d{\bf r}\right )^{N},
\label{lo4}
\end{equation}
where ${\bf f}$ is given by Eq. (\ref{in1}). Using $\int_{|{\bf r}|=0}^{R} \tau({\bf r})\, d{\bf r}=1$, and $\tau({\bf r})=n({\bf r})/N$, the foregoing expression is equivalent to
\begin{equation}
A_{N}({\bf k})=\left (1-\frac{1}{N}\int_{|{\bf r}|=0}^{R} \left (1-e^{i {\bf k}\cdot {\bf f}}\right ) n({\bf r}) \, d{\bf r}\right )^{N}.
\label{lo5}
\end{equation}
We now consider the  limit $N\rightarrow +\infty$, $R\rightarrow +\infty$ with $N/R^{d-p}$ fixed. In this limit, the distribution of the force can be written
\begin{equation}
W({\bf F})=\frac{1}{(2\pi)^{d}}\int A({\bf k})e^{-i {\bf k}\cdot {\bf
F}}d{\bf k},
\label{df1}
\end{equation}
with
\begin{eqnarray}
A({\bf k})=e^{-C({\bf k})}, \qquad C({\bf k})=\int_{|{\bf r}|=0}^{+\infty}
n({\bf r}) \left (1-e^{i {\bf k}\cdot {\bf f}}\right )d{\bf r}.\nonumber\\
\label{df2}
\end{eqnarray}
Using the transformation (\ref{df3}), we obtain
\begin{eqnarray}
C({\bf k})=\frac{K}{d+\alpha-1}(Gm)^{\frac{d-p}{d+\alpha-1}}\nonumber\\
\times\int_{|{\bf f}|=0}^{+\infty} \left (1-e^{i {\bf k}\cdot {\bf f}}\right ) f^{\frac{p-d(d+\alpha)}{d+\alpha-1}}\, d{\bf f}.
\label{df4}
\end{eqnarray}
The characteristic function $C({\bf k})$ converges for $f\rightarrow
+\infty$ if $p<d$ and $d+\alpha-1>0$, and for $f\rightarrow 0$ if
$2\alpha+p>2-d$. In other words, it converges if
$d<\gamma<\gamma_c$. Introducing a spherical system of coordinates,
we get
\begin{eqnarray}
C({\bf k})=\frac{KC_{d}}{d+\alpha-1}(Gm)^{\frac{d-p}{d+\alpha-1}}\int_{0}^{+\infty}df \ f^{d-1}\nonumber\\
\times\int_{0}^{\pi}d\theta\  (\sin\theta)^{d-2}   \left (1-\cos(kf\cos\theta)\right ) f^{\frac{p-d(d+\alpha)}{d+\alpha-1}}.
\label{df5}
\end{eqnarray}
We have introduced the notation $C_{d}=S_{d}/\int_{0}^{\pi}  (\sin\theta)^{d-2}\, d\theta$ where $S_{d}=2\pi^{d/2}/\Gamma(d/2)$ represents the surface of a unit
sphere in $d$ dimensions. Using the
identity
\begin{eqnarray}
\int_{0}^{\pi}  (\sin\theta)^{d-2}\, d\theta=\frac{\sqrt{\pi}\Gamma\left (\frac{d-1}{2}\right )}{\Gamma\left (\frac{d}{2}\right )},
\label{df8}
\end{eqnarray}
we obtain
\begin{eqnarray}
C_{d}=\frac{2\pi^{\frac{d-1}{2}}}{\Gamma\left (\frac{d-1}{2}\right )}.
\label{df6}
\end{eqnarray}
Next, setting $x=kf$ and using the
identity
\begin{eqnarray}
\int_{0}^{\pi} \cos(x\cos\theta) (\sin\theta)^{d-2}  d\theta\nonumber\\
=\sqrt{\pi}\left (\frac{2}{x}\right )^{\frac{d}{2}-1} J_{\frac{d}{2}-1}(x)\Gamma\left (\frac{d-1}{2}\right ),
\label{df7}
\end{eqnarray}
we can rewrite Eq. (\ref{df5}) in the form
\begin{eqnarray}
C({\bf k})=a k^{H}, 
\label{df9}
\end{eqnarray}
where 
\begin{eqnarray}
H=\frac{d-p}{d+\alpha-1}=\gamma-d, 
\label{df10}
\end{eqnarray}
\begin{eqnarray}
a=\frac{S_{d}}{d+\alpha-1}(Gm)^{H}KB,
\label{df11}
\end{eqnarray}
\begin{eqnarray}
B=\int_{0}^{+\infty}\frac{dx}{x^{H+1}}\left \lbrack 1-\Gamma\left (\frac{d}{2}\right )\left (\frac{2}{x}\right )^{\frac{d}{2}-1}J_{\frac{d}{2}-1}(x)\right \rbrack.\nonumber\\
\label{df12}
\end{eqnarray}
The integral converges for $x\rightarrow +\infty$ if $H>0$, i.e. if $d>p$ and $d+\alpha-1>0$ or, equivalently, if $\gamma>d$. On the other hand, using
\begin{eqnarray}
J_{\nu}(x)\sim_{0} \frac{1}{\Gamma(\nu+1)}\left (\frac{x}{2}\right )^{\nu}-\frac{1}{\Gamma(\nu+2)}\left (\frac{x}{2}\right )^{\nu+2},
\label{df13}
\end{eqnarray}
we see that the integral converges for $x\rightarrow 0$ if $H<2$, i.e. if $2\alpha+p>2-d$ or, equivalently, if $\gamma<\gamma_{c}$. Making an integration by parts and using the identity
\begin{eqnarray}
\frac{d}{dx}\left \lbrack \frac{J_{n}(x)}{x^{n}}\right\rbrack=\frac{J_{n}'(x)-\frac{n}{x}J_{n}(x)}{x^{n}}=-\frac{J_{n+1}(x)}{x^{n}},
\label{df14}
\end{eqnarray}
we can rewrite the function $B$ in the form 
\begin{eqnarray}
B=\frac{1}{H}\Gamma\left (\frac{d}{2}\right )2^{\frac{d}{2}-1}\int_{0}^{+\infty} J_{\frac{d}{2}}(x)\frac{dx}{x^{H+\frac{d}{2}-1}}.
\label{df15}
\end{eqnarray}
It can also be expressed in terms of Gamma functions as 
\begin{eqnarray}
B=\frac{1}{2^H}\frac{1}{H}\Gamma\left (\frac{d}{2}\right )\frac{\Gamma\left (1-\frac{H}{2}\right )}{\Gamma\left (\frac{d+H}{2}\right )}. 
\label{tu1}
\end{eqnarray}
The distribution of the force is given by Eq. (\ref{df1}). Introducing a spherical system of coordinates, it can be rewritten
\begin{eqnarray}
W({\bf F})=\frac{C_d}{(2\pi)^{d}}\int_{0}^{+\infty}dk\, k^{d-1}\nonumber\\ 
\times\int_{0}^{\pi}d\theta \, (\sin\theta)^{d-2}e^{-a k^{H}}e^{-ikF\cos\theta}.
\label{tu2}
\end{eqnarray}
Using identity (\ref{df7}), we obtain
\begin{eqnarray}
W({\bf F})=\frac{S_{d}}{(2\pi)^{d}}\Gamma\left (\frac{d}{2}\right )\left (\frac{2}{F}\right )^{\frac{d}{2}-1}\nonumber\\
\times\int_{0}^{+\infty} e^{-ak^{H}}k^{\frac{d}{2}}J_{\frac{d}{2}-1}(kF)dk.
\label{df16}
\end{eqnarray}
Note that the structure of the distribution only depends on the
scaling exponent $H$ which takes values in the range
\begin{eqnarray}
0<H<H_c=2.
\label{df16b}
\end{eqnarray} 
The asymptotic behaviour of $W({\bf F})$ for small $|{\bf F}|$ can be obtained by expanding the Bessel function in Taylor series and integrating term by term. This yields
\begin{eqnarray}
W({\bf F})=\frac{S_{d}}{(2\pi)^{d}H}\Gamma\left (\frac{d}{2}\right
)\sum_{l=0}^{+\infty}\frac{(-1)^{l}}{l! 4^{l}}\frac{\Gamma\left (\frac{2l+d}{H}\right )}{\Gamma\left
(\frac{d}{2}+l\right )}\frac{F^{2l}}{a^{\frac{2l+d}{H}}}.\nonumber\\
\label{tue1}
\end{eqnarray} 
The asymptotic behaviour of $W({\bf F})$ for large $|{\bf F}|$ can be obtained by using a method similar to the one exposed in Sec. II.C. of \cite{cs1}. Performing the changes of variables  $z=kF$ and $t=-\cos\theta$ in Eq. (\ref{tu2}), we obtain 
\begin{eqnarray}
W({\bf
F})=\frac{C_d}{(2\pi)^{d}F^{d}}\int_{-1}^{+1}(1-t^2)^{\frac{d-3}{2}}\,
dt \nonumber\\
\times\int_{0}^{+\infty}e^{-a \left (\frac{z}{F}\right )^{H}}e^{izt}z^{d-1}\, dz.
\label{tu3}
\end{eqnarray} 
In this expression, $t$ and $z$ are real and the domains of
integration are on the real axis $-1\le t\le 1$ and $z\ge 0$. Under
this form, we cannot expand the exponential in power series for
$F\rightarrow +\infty$ and integrate term by term because the
integrals would diverge. The idea is to work in the complex plane
and deform the contours of integration as indicated in Sec. II.C. of
\cite{cs1}. It is then possible to perform the integration on $t$
along the semi-circle $C_{+}$ of radius unity in the upper-half plane
${\rm Im}(t)\ge 0$, and the integration on $z$ along the line such
that $izt=-y$ with $y$ real $\ge 0$. We then obtain
\begin{eqnarray}
W({\bf F})=\frac{C_d}{(2\pi)^{d}F^{d}}{\rm Re}\int_{C_{+}} (1-t^2)^{\frac{d-3}{2}}\, dt \nonumber\\
\times\int_{0}^{+\infty}e^{-a \left (\frac{iy}{tF}\right )^{H}}e^{-y}\left (\frac{i}{t}\right )^{d}y^{d-1}\, dy.
\label{tu4}
\end{eqnarray} 
We can now expand the exponential term in Taylor series and perform the integration on $y$ to obtain
\begin{eqnarray}
W({\bf F})=\frac{C_d}{(2\pi)^{d}}\sum_{l=0}^{+\infty}{\rm Re}\int_{C_{+}} \frac{(1-t^2)^{\frac{d-3}{2}}}{t^{Hl+d}}\nonumber\\
\times\frac{(-a)^{l}}{l!}\frac{i^{Hl+d}}{F^{Hl+d}}\Gamma(Hl+d). 
\label{tu5}
\end{eqnarray} 
We now need to evaluate the integral
\begin{eqnarray}
I={\rm Re}\int_{C_{+}} \frac{(1-t^2)^{\frac{d-3}{2}}}{t^{Hl+d}}i^{Hl+d},
\label{tu6}
\end{eqnarray} 
with $t=e^{i\theta}$ and $\theta$ going from $\pi$ to $0$. After straightforwad calculations, we can rewrite the integral (\ref{tu6}) in the form
\begin{eqnarray}
I=-2^{\frac{d-3}{2}}\int_{0}^{\pi}\cos \biggl\lbrack \left (\frac{1}{2}+\frac{d}{2}+Hl\right )\theta\nonumber\\
-\left (Hl+\frac{d}{2}+\frac{5}{2}\right )\frac{\pi}{2}\biggr\rbrack (\sin\theta)^{\frac{d-3}{2}}\, d\theta.
\label{tu7}
\end{eqnarray} 
Using the identity
\begin{eqnarray}
\int_{0}^{\pi}\cos  (a\theta+b) (\sin \theta)^{k}\, d\theta\nonumber\\
=\frac{\pi}{2^k}\frac{\Gamma(1+k)}{\Gamma\left (1-\frac{a}{2}+\frac{k}{2}\right )\Gamma\left (1+\frac{a}{2}+\frac{k}{2}\right )}\cos\left ( a\frac{\pi}{2}+b\right ),\nonumber\\
\label{tu8}
\end{eqnarray} 
we find that
\begin{eqnarray}
I=\frac{\pi \Gamma\left (\frac{d-1}{2}\right )}{\Gamma\left (-\frac{Hl}{2}\right )\Gamma\left (\frac{1+d+Hl}{2}\right )}.
\label{tu9}
\end{eqnarray} 
Combining the previous results, the large $F$ expansion of the distribution of the force can be written
\begin{eqnarray}
W({\bf F})=\frac{C_d}{(2\pi)^{d}}\sum_{l=0}^{+\infty} \frac{(-a)^{l}}{l!}\frac{1}{F^{Hl+d}}\Gamma(Hl+d)\nonumber\\
\times\frac{\pi \Gamma\left (\frac{d-1}{2}\right )}{\Gamma\left (-\frac{Hl}{2}\right )\Gamma\left (\frac{1+d+Hl}{2}\right )}.
\label{tu10}
\end{eqnarray} 
We then obtain the equivalent for $F\rightarrow +\infty$:
\begin{eqnarray}
W({\bf F})\sim -\frac{C_d}{(2\pi)^{d}} \frac{a}{F^{H+d}}\Gamma(H+d)\nonumber\\
\times\frac{\pi \Gamma\left (\frac{d-1}{2}\right )}{\Gamma\left (-\frac{H}{2}\right )\Gamma\left (\frac{1+d+H}{2}\right )}.
\label{tu11}
\end{eqnarray} 
Now, combining Eqs. (\ref{df11}) and (\ref{tu1}), we have
\begin{eqnarray}
a=-\frac{\pi^{d/2}}{d+\alpha-1}(Gm)^{H}K\frac{1}{2^{H}}\frac{\Gamma\left (-\frac{H}{2}\right )}{\Gamma\left (\frac{d+H}{2}\right )}.
\label{tu12}
\end{eqnarray} 
Substituting this expression in Eq. (\ref{tu11}), using Eq. (\ref{df6}) and the identity 
\begin{eqnarray}
2^{2z-1}\Gamma(z)\Gamma\left (z+\frac{1}{2}\right )=\sqrt{\pi}\Gamma(2z),
\label{tu13}
\end{eqnarray} 
we finally obtain
\begin{equation}
W({\bf F})\sim K\frac{(Gm)^{H}}{d+\alpha-1}\frac{1}{F^{d+H}}\qquad (F\rightarrow +\infty).
\label{df17}
\end{equation}
Therefore, the tail of the distribution decreases with the exponent
\begin{eqnarray}
\gamma=d+H,
\label{df18}
\end{eqnarray}
like for the individual distribution (\ref{mon1}). Furthermore, we
shall see in Sec. \ref{sec_nnw} that the asymptotic behaviour of the
distribution of the force coincides with the expression (\ref{nnw6})
derived in the nearest neighbor approximation.

The previous results can also be used for analyzing the stochastic
gravitational fluctuations generated by a fractal distribution of
field sources (stars or galaxies) provided that we make the
correspondance
\begin{equation}
p=d-d_{f},
\label{nnw9}
\end{equation}
where $d_{f}$ is the fractal dimension of the distribution in a
$d$-dimensional universe \cite{vald}.  We also introduce the exponent
$\nu=d+\alpha-1$ characterizing the power-law decay of the force. For
the gravitational interaction $\nu=d-1$.  In terms of these
quantities, the scaling exponent $H$ can be written
\begin{equation}
H=\frac{d_{f}}{\nu}.
\label{nnw11}
\end{equation} 
The condition (\ref{df16b}) required to have a L\'evy law is
\begin{equation}
0< d_{f}< 2\nu.
\label{nnw10}
\end{equation}
Such a formalism can be useful in cosmology where observations suggest
that galaxies are distributed according to a fractal law characterized
by a fractal dimension $1<d_{f}<2$ (in a $d=3$ universe) \cite{cp,mandel}. For this
range of values, the scaling exponent satisfies $1/2<H<1$.

\subsection{Nearest neighbor approximation}
\label{sec_nnw}

Let us compare these results with those obtained by making the nearest
neighbor approximation (see also Appendix \ref{sec_pvg}).  For an
arbitrary inhomogeneous distribution of particles, the distribution of
the nearest neighbor is obtained from the relation
\begin{equation}
\tau_{n.n.}(r)dr=\left (1-\int_{0}^{r}\tau_{n.n.}(r')dr'\right )n({\bf r})S_{d}r^{d-1}dr,
\label{nnw1}
\end{equation}
leading to
\begin{equation}
\tau_{n.n.}({\bf r})=\frac{\tau_{n.n.}({r})}{S_{d}r^{d-1}}=n({\bf r})e^{-\int_{0}^{r} n({\bf x}) S_{d} x^{d-1} dx}.
\label{nnw2}
\end{equation}
For a power-law density profile $n({\bf r})=K/r^{p}$, we get
\begin{equation}
\tau_{n.n.}({\bf r})=\frac{K}{r^{p}}e^{-\frac{S_{d}K r^{d-p}}{d-p}}.
\label{nnw3}
\end{equation}
The distribution of the force due to the nearest neighbor is obtained from
the relation $W_{n.n.}({\bf F})d{\bf F}=\tau_{n.n.}({\bf r})d{\bf r}$ with
\begin{equation}
{\bf F}=Gm\frac{{\bf r}}{r^{(d+\alpha)}}.
\label{tue2}
\end{equation}
Using
\begin{equation}
d{\bf F}=(d+\alpha-1)(Gm)^{-\frac{d}{d+\alpha-1}}F^{\frac{d(d+\alpha)}{d+\alpha-1}}\, d{\bf r},
\label{nnw4}
\end{equation}
we obtain
\begin{eqnarray}
W_{n.n.}({\bf F})=K\frac{(Gm)^{H}}{d+\alpha-1}F^{-(d+H)}
 e^{-\frac{S_{d}K}{d-p}\left (\frac{Gm}{F}\right )^{H}}.
\label{nnw5}
\end{eqnarray}
For $F\rightarrow +\infty$, we get the asymptotic behaviour
\begin{equation}
W_{n.n.}({\bf F})\sim K\frac{(Gm)^{H}}{d+\alpha-1}F^{-(d+H)},
\label{nnw6}
\end{equation}
which coincides with the asymptotic behaviour 
(\ref{df17}) of the exact distribution. Therefore, the tail of the
distribution is dominated by the contribution of the nearest
neighbour.  We note that the moment of order $b$ of the distribution
(\ref{nnw6}) is finite for $b<H$ and its value is
\begin{equation}
\langle F^b\rangle_{n.n.}=\left (\frac{S_d K}{d-p}\right )^{b/H}(Gm)^{b} \Gamma\left (1-\frac{b}{H}\right ).
\label{nnw8}
\end{equation}

\subsection{The dimension $d=3$}
\label{sec_trois}

For the ordinary dimension $d=3$, we have
\begin{eqnarray}
C({\bf k})=a k^{H}, 
\label{t1}
\end{eqnarray}
where 
\begin{equation}
H=\frac{3-p}{2+\alpha},
\label{t2}
\end{equation}
\begin{equation}
a=\frac{4\pi}{2+\alpha}(Gm)^{H}KB,
\label{t3}
\end{equation}
\begin{equation}
K=\frac{(3-p)N}{4\pi R^{3-p}}.
\label{t4}
\end{equation}
As before, we assume that $0\le p<3$, $\alpha>-2$ and $2\alpha+p>-1$
(or, equivalently, $3<\gamma<5$). Therefore, $0<H<2$. On the other
hand, using
\begin{equation}
J_{1/2}(x)=\sqrt{\frac{2}{\pi x}}\sin x,
\label{t5}
\end{equation}
and $\Gamma(3/2)=\sqrt{\pi}/2$, the function $B$ defined by
Eq. (\ref{df12}) takes the form
\begin{eqnarray}
B=\int_{0}^{+\infty}\frac{dx}{x^{H+2}}(x-\sin x)=\frac{1}{H(H+1)}\int_{0}^{+\infty}\frac{\sin x}{x^{H}}dx,\nonumber\\
\label{t6}
\end{eqnarray}
where we have used two integrations by parts to get the last equality. This expression can also be obtained from Eq. (\ref{df15}) by using the identity 
\begin{equation}
J_{3/2}(x)=\sqrt{\frac{2}{\pi x}}\left (\frac{\sin x}{x}-\cos x\right ),
\label{t7}
\end{equation}
and performing an integration by parts. The function $B$ can finally be expressed
in terms of the Gamma function as
\begin{equation}
B=\frac{\pi}{2\Gamma(H+2)\sin\left (\frac{H\pi}{2}\right )}.
\label{t8}
\end{equation}
Using identity (\ref{tu13}) and the identity
\begin{equation}
\Gamma(x)\Gamma(1-x)=\frac{\pi}{\sin (\pi x)},
\label{t16nd}
\end{equation} 
we can check that Eq. (\ref{t8}) is consistent with Eq. (\ref{tu1}).
The distribution of the force is then given by
\begin{equation}
W({\bf F})=\frac{1}{2\pi^{2}F}\int_{0}^{+\infty}e^{-ak^{H}} \sin(kF) k\, dk,
\label{t9}
\end{equation}
where we recall that $0<H<2$. For $\alpha=0$ (gravity), we recover the
situation considered by Kandrup \cite{kandrup}. In that case,
$H=(3-p)/2$. If in addition $p=0$ (homogeneous system), we recover the
situation considered by Chandrasekhar \cite{chandra}. In that case,
$H=3/2$, $K=n$, $B=4\sqrt{2\pi}/15$ and $a=\frac{4}{15}(2\pi
Gm)^{3/2}n$. We will see that certain results derived by Kandrup
\cite{kandrup} contain some mistakes, so it is important to reconsider
this situation in detail. In addition, we treat the case of a general
power-law potential where $\alpha$ can be non-zero.

The general properties of the distribution (\ref{t9}) have been
derived by Chandrasekhar \cite{c48}. Although Chandrasekhar considered
a uniform medium, it is important to note that his results remain
valid for a power-law density profile; we just need to replace the
index that appears in his analysis by $H$.  Let us rewrite the results
obtained by Chandrasekhar \cite{c48} with the present notations. When
$F$ is small, a convenient series expansion of the distribution is
given by \cite{c48}: 
\begin{equation}
W({\bf F})=\frac{1}{2\pi^{2}H}\sum_{l=0}^{+\infty}(-1)^{l}\Gamma\left (\frac{2l+3}{H}\right )\frac{1}{a^{\frac{2l+3}{H}}}\frac{F^{2l}}{(2l+1)!}.
\label{t10}
\end{equation}
In particular,
\begin{equation}
W({\bf F})\rightarrow \frac{1}{2\pi^{2}H}\Gamma\left (\frac{3}{H}\right )\frac{1}{a^{{3}/{H}}}, \qquad (F\rightarrow 0).
\label{t11}
\end{equation}
Using identity (\ref{tu13}), we can check that Eq. (\ref{t10}) is consistent with Eq. (\ref{tue1}). For $F\rightarrow +\infty$, we have the series expansion \cite{c48}: 
\begin{eqnarray}
W({\bf F})=\frac{1}{2\pi^{2}}\sum_{l=1}^{+\infty}(-1)^{l+1}\frac{a^{l}}{l!}\Gamma (Hl+2)\sin\left (Hl\frac{\pi}{2}\right )\frac{1}{F^{3+lH}}.\nonumber\\
\label{t12}
\end{eqnarray}
In particular,
\begin{eqnarray}
W({\bf F})\sim \frac{1}{2\pi^{2}} a \Gamma (H+2)\sin\left (H\frac{\pi}{2}\right )\frac{1}{F^{3+H}}, \qquad (F\rightarrow +\infty).\nonumber\\
\label{t13}
\end{eqnarray}
Using Eqs. (\ref{t3}) and (\ref{t8}), we have
\begin{equation}
W({\bf F})\sim \frac{K(Gm)^{H}}{2+\alpha} \frac{1}{F^{3+H}}, \qquad (F\rightarrow +\infty).
\label{t14}
\end{equation}
This asymptotic behaviour coincides\footnote{Note that the expression
of the asymptotic behaviour of the distribution of the force obtained
by Kandrup \cite{kandrup} is not correct. Thus, his conclusion that
the asymptotic distribution of the force does not {\it exactly}
coincide with the distribution produced by the nearset neighbor must
be revised.} with the asymptotic behaviour of the distribution of the
force (\ref{nnw6}) due to the nearest neighbor. Using identity (\ref{tu13}), we can check that Eqs. (\ref{t12})-(\ref{t14}) are consistent with Eqs. 
(\ref{tu10}), (\ref{tu11}) and (\ref{df17}). The moments of the
force are finite iff $-3<b<H$. For $0\le b<H$, we have \cite{c48}:
\begin{equation}
\langle F^{b}\rangle=a^{b/H}\frac{2}{\pi}(b+1)\Gamma(b)\Gamma\left (1-\frac{b}{H}\right )\sin\left (\frac{b\pi}{2}\right ).
\label{t15}
\end{equation}
In particular, for $b=1$, we get
\begin{equation}
\langle F\rangle=a^{1/H}\frac{4}{\pi}\Gamma\left (1-\frac{1}{H}\right ).
\label{t16}
\end{equation}
On the other hand, if we start directly from Eq. (39) of Chandrasekhar
\cite{c48} and use the identity
\begin{equation}
\int_{0}^{+\infty}\frac{\sin x}{x^{\alpha}}\, dx=\frac{\pi}{2\Gamma(\alpha)\sin \left (\frac{\alpha\pi}{2}\right )},\qquad (0<\alpha<2).
\label{u25}
\end{equation}
we find that for $-1<b<{\rm min}(1,H)$:
\begin{equation}
\langle F^{b}\rangle=a^{b/H}(b+1)\Gamma\left (1-\frac{b}{H}\right )\frac{1}{\Gamma(1-b) \cos\left (\frac{b\pi}{2}\right )}.
\label{t15nd}
\end{equation}
The link with expression (\ref{t15}) in the common interval of $b$ is
made by using the identity (\ref{t16nd}).  Finally, if we start
directly from Eq. (24) of Chandrasekhar \cite{c48} and use the
identity (\ref{u25}), we find that for $-3<b<-1$:
\begin{equation}
\langle F^{b}\rangle=-\frac{1}{H}a^{b/H}\frac{1}{\Gamma (-b-1) \cos\left (\frac{b\pi}{2}\right )}\Gamma\left (-\frac{b}{H}\right ).
\label{t17nd}
\end{equation}
Finally, using the identity $\Gamma(x+1)=x\Gamma(x)$, one easily checks
that Eqs. (\ref{t15}), (\ref{t15nd}) and (\ref{t17nd}) coincide in the
whole range $0\le b<H$.

\subsection{The dimension $d=1$}
\label{sec_un}

For the dimension $d=1$, we have
\begin{eqnarray}
C({\bf k})=a k^{H}, 
\label{u1}
\end{eqnarray}
where 
\begin{equation}
H=\frac{1-p}{\alpha},
\label{u2}
\end{equation}
\begin{equation}
a=\frac{2}{\alpha}(Gm)^{H}KB,
\label{u3}
\end{equation}
\begin{equation}
K=\frac{(1-p)N}{2 R^{1-p}}.
\label{u4}
\end{equation}
As before, we assume that $0\le p<1$, $\alpha>0$ and $2\alpha+p>1$ (or, equivalently, $1<\gamma<3$). Therefore, $0<H<2$. On the other hand,
using
\begin{eqnarray}
J_{-1/2}(x)=\sqrt{\frac{2}{\pi x}}\cos x,
\label{u5}
\end{eqnarray}
and $\Gamma(1/2)=\sqrt{\pi}$, the function $B$ defined by
Eq. (\ref{df12}) takes the form
\begin{eqnarray}
B=\int_{0}^{+\infty}\frac{dx}{x^{H+1}}(1-\cos x)=\frac{1}{H}\int_{0}^{+\infty}\frac{\sin x}{x^{H}}dx,\nonumber\\
\label{u6}
\end{eqnarray}
where we have used an integration by parts to get the last equality. This expression can also be obtained from Eq. (\ref{df15}) by using the identity 
\begin{equation}
J_{1/2}(x)=\sqrt{\frac{2}{\pi x}}\sin x.
\label{u7}
\end{equation}
The function $B$ can finally be expressed
in terms of the Gamma function as
\begin{equation}
B=\frac{\pi}{2\Gamma(H+1)\sin\left (\frac{H\pi}{2}\right )}.
\label{u8}
\end{equation}
Using identities (\ref{tu13}) and (\ref{t16nd}), we can
check that Eq. (\ref{u8}) is consistent with Eq. (\ref{tu1}). The
distribution of the force is then given by
\begin{equation}
W({F})=\frac{1}{\pi}\int_{0}^{+\infty}e^{-ak^{H}} \cos(kF) \, dk,
\label{u9}
\end{equation}
where we recall that $0<H<2$. The general properties of the
distribution (\ref{u9}) can be derived by adapting the method
developed by Chandrasekhar \cite{c48} for $d=3$.  When $F$ is small, a
convenient series expansion of $W(F)$ is obtained by expanding
$\cos(kF)$ in Taylor series and integrating term by term. In this
manner, we obtain
\begin{equation}
W({F})=\frac{1}{\pi H}\sum_{l=0}^{+\infty}(-1)^{l}\Gamma\left (\frac{2l+1}{H}\right )\frac{1}{a^{\frac{2l+1}{H}}}\frac{F^{2l}}{(2l)!}.
\label{u10}
\end{equation}
In particular,
\begin{equation}
W({F})\rightarrow \frac{1}{\pi H}\Gamma\left (\frac{1}{H}\right )\frac{1}{a^{{1}/{H}}}, \qquad (F\rightarrow 0).
\label{u11}
\end{equation}
Using identity (\ref{tu13}), we can check that Eq. (\ref{u10}) is
consistent with Eq. (\ref{tue1}).  To obtain the series expansion of
$W(F)$ for $F\rightarrow +\infty$, we first set $z=kF$ and rewrite
Eq. (\ref{u9}) in the form
\begin{equation}
W({F})=\frac{1}{\pi F}{\rm Re} \int_{0}^{+\infty}e^{-a \left (\frac{z}{F}\right )^{H}} e^{iz} \, dz.
\label{u12}
\end{equation}
We now integrate in the complex plane along the line passing through the origin and inclined at an angle $\frac{\pi}{2H}$ to the real axis instead of along the real axis itself. Thus, we set $z=e^{i\frac{\pi}{2H}}y$ where $y$ is real $\ge 0$ in Eq. (\ref{u12}) and we obtain
\begin{eqnarray}
W({F})=\frac{1}{\pi F}{\rm Re} \ e^{i\frac{\pi}{2H}}\int_{0}^{+\infty}e^{-i a \left (\frac{y}{F}\right )^{H}}\nonumber\\
\times {\rm exp}\left\lbrace -e^{-i(1-\frac{1}{H})\frac{\pi}{2}}y\right\rbrace  \, dy.
\label{u13}
\end{eqnarray}
Expanding $e^{-i a \left (\frac{y}{F}\right )^{H}}$ in Taylor series, we have
\begin{eqnarray}
W({F})=\frac{1}{\pi F}{\rm Re} \ \sum_{l=0}^{+\infty} (-i)^{l} \frac{a^{l}}{l!} e^{i\frac{\pi}{2H}} \frac{1}{F^{Hl}}\nonumber\\
\times \int_{0}^{+\infty} {\rm exp}\left\lbrace -e^{-i(1-\frac{1}{H})\frac{\pi}{2}}y\right\rbrace y^{Hl} \, dy.
\label{u14}
\end{eqnarray}
We rotate again the line of integration by an angle $(1-\frac{1}{H})\frac{\pi}{2}$. Thus, we set $z=e^{-i(1-\frac{1}{H})\frac{\pi}{2}}y$ where $z$ is real $\ge 0$ and we obtain
\begin{eqnarray}
W({F})=\frac{1}{\pi F}{\rm Re} \ \sum_{l=0}^{+\infty} (-i)^{l} \frac{a^{l}}{l!} e^{i\frac{\pi}{2H}} \frac{1}{F^{Hl}}\nonumber\\
\times e^{i(Hl+1)(1-\frac{1}{H})\frac{\pi}{2}}\Gamma(Hl+1).
\label{u15}
\end{eqnarray}
Now, we verify that
\begin{eqnarray}
{\rm Re} \ (-i)^{l}  e^{i\frac{\pi}{2H}} e^{i(Hl+1)(1-\frac{1}{H})\frac{\pi}{2}}=(-1)^{l+1}\sin\left (\frac{Hl\pi}{2}\right ).\nonumber\\
\label{u16}
\end{eqnarray}
Hence,  we obtain the asymptotic expansion for $F\rightarrow +\infty$:
\begin{eqnarray}
W({F})=\frac{1}{\pi}\sum_{l=1}^{+\infty}(-1)^{l+1}\frac{a^{l}}{l!}\Gamma (Hl+1)\sin\left (Hl\frac{\pi}{2}\right )\frac{1}{F^{1+lH}}.\nonumber\\
\label{u17}
\end{eqnarray}
In particular,
\begin{eqnarray}
W({F})\sim \frac{1}{\pi} a \Gamma (H+1)\sin\left (H\frac{\pi}{2}\right )\frac{1}{F^{1+H}}, \qquad (F\rightarrow +\infty).\nonumber\\
\label{u18}
\end{eqnarray}
Using Eqs. (\ref{u3}) and (\ref{u8}), this can be rewritten
\begin{equation}
W({F})\sim \frac{K(Gm)^{H}}{\alpha} \frac{1}{F^{1+H}}, \qquad (F\rightarrow +\infty).
\label{u19}
\end{equation}
This asymptotic behaviour coincides with the asymptotic behaviour of
the distribution of the force (\ref{nnw6}) due to the nearest
neighbor.

To evaluate the asymptotic expansion of $W(F)$ given by
Eq. (\ref{u12}) for $F\rightarrow +\infty$, we can also integrate along
the imaginary axis. Thus, we set $z=iy$ with $y$ real $\ge 0$. In that
case, Eq. (\ref{u12}) becomes
\begin{eqnarray}
W({F})=\frac{1}{\pi F}{\rm Re} \ i \int_{0}^{+\infty}e^{- a \left (\frac{iy}{F}\right )^{H}}e^{-y}\, dy.
\label{u20}
\end{eqnarray}
Expanding the first term in the integral in Taylor series, we get
\begin{eqnarray}
W({F})=\frac{1}{\pi F}{\rm Re} \ \sum_{l=0}^{+\infty} i^{Hl+1} \frac{(-a)^{l}}{l!} \frac{1}{F^{Hl}}\Gamma(Hl+1).
\label{u21}
\end{eqnarray}
Noting that
\begin{eqnarray}
{\rm Re} \ i^{Hl+1} =-\sin\left (\frac{Hl\pi}{2}\right ),\nonumber\\
\label{u22}
\end{eqnarray}
we recover Eq. (\ref{u17}). 

From the asymptotic behaviour (\ref{u19}), it is clear that the
moments of the force $\langle
F^b\rangle=2\int_{0}^{+\infty}W(F)F^{b}\, dF$ are finite iff $-1<
b<H$. We can obtain an analytical expression for $-1<b<0$. For ease of
notations, we set $\nu=-b$ with $0<\nu<1$. From Eq. (\ref{u9}), we
have
\begin{eqnarray}
\langle F^{-\nu} \rangle=\frac{2}{\pi}\int_{0}^{+\infty}\int_{0}^{+\infty} e^{-ak^{H}}F^{-\nu} \cos (kF)\, dF dk.\nonumber\\
\label{u24}
\end{eqnarray}
We first integrate on $F$ using the identity 
\begin{equation}
\int_{0}^{+\infty}\frac{\cos x}{x^{\alpha}}\, dx=\frac{\pi}{2\Gamma(\alpha)\cos \left (\frac{\alpha\pi}{2}\right )},\qquad (0<\alpha<1).
\label{u25mod}
\end{equation}
This yields
\begin{eqnarray}
\langle F^{-\nu} \rangle 
=\frac{1}{\Gamma(\nu)\cos (\nu\frac{\pi}{2})}\int_{0}^{+\infty} e^{-ak^{H}} k^{\nu-1}\, dk.
\label{u27}
\end{eqnarray}
Expressing the integral in terms of $\Gamma$-functions, we finally obtain the formula
\begin{eqnarray}
\langle F^{-\nu}\rangle = \frac{\Gamma(\nu/H)}{H\Gamma(\nu)\cos (\nu\frac{\pi}{2})a^{\nu/H}}, \quad (0<\nu<1).
\label{u28}
\end{eqnarray}

\subsection{The dimension $d=2$}
\label{sec_de}

For the dimension $d=2$, we have
\begin{eqnarray}
C({\bf k})=a k^{H}, 
\label{de1}
\end{eqnarray}
where 
\begin{eqnarray}
H=\frac{2-p}{1+\alpha},
\label{de2}
\end{eqnarray}
\begin{eqnarray}
a=\frac{2\pi}{1+\alpha}(Gm)^{H}KB,
\label{de3}
\end{eqnarray}
\begin{equation}
K=\frac{(2-p)N}{2\pi R^{2-p}}.
\label{de4}
\end{equation}
As before, we assume that $0\le p<2$, $\alpha>-1$ and $2\alpha+p>0$ (or, equivalently, $2<\gamma<4$). Therefore, $0<H<2$. The function $B$ is given by 
\begin{eqnarray}
B=\int_{0}^{+\infty}\frac{dx}{x^{H+1}}\left \lbrack 1-J_{0}(x)\right \rbrack.
\label{de5}
\end{eqnarray}
Integrating by parts, the foregoing integral can be rewritten
\begin{eqnarray}
B=\frac{1}{H}\int_{0}^{+\infty}\frac{dx}{x^{H}}J_{1}(x).
\label{de6}
\end{eqnarray}
Finally, the function $B$ can be expressed in terms of $\Gamma$-functions under the form 
\begin{eqnarray}
B=\frac{1}{2^H}\frac{1}{H}\frac{\Gamma \left (1-\frac{H}{2}\right )}{\Gamma \left (1+\frac{H}{2}\right )},
\label{de7}
\end{eqnarray}
in agreement with Eq. (\ref{tu1}).  The distribution of the force is
then given by
\begin{eqnarray}
W({\bf F})=\frac{1}{2\pi}\int_{0}^{+\infty} e^{-ak^{H}} J_{0}(kF)k\, dk,
\label{de8}
\end{eqnarray}
where we recall that $0<H<2$. When $F$ is small, a convenient series
expansion of $W({\bf F})$ is obtained by expanding $J_{0}(kF)$ is
Taylor series and integrating term by term. This yields
\begin{eqnarray}
W({\bf F})=\frac{1}{2\pi H}\sum_{l=0}^{+\infty}\frac{(-1)^{l}}{(l!)^{2}}\frac{F^{2l}}{4^{l}}\frac{1}{a^{\frac{2l+2}{H}}}\Gamma\left (\frac{2l+2}{H}\right ),
\label{de9}
\end{eqnarray}
in agreement with Eq. (\ref{tue1}). The asymptotic
expansion of the distribution $W({\bf F})$ for large $F$ can be
obtained from the general method developed in Sec. \ref{sec_df}
leading to
\begin{eqnarray}
W({\bf F})=\frac{1}{2\sqrt{\pi}}\sum_{l=0}^{+\infty} \frac{(-a)^{l}}{l!}\frac{1}{F^{Hl+2}}\frac{\Gamma(Hl+2)}{\Gamma\left (-\frac{Hl}{2}\right )\Gamma\left (\frac{3+Hl}{2}\right )}.\nonumber\\
\label{gjrwe}
\end{eqnarray} 
Using Eqs. (\ref{de3}), (\ref{de7}) and the identity (\ref{tu13}), we obtain the equivalent for $F\rightarrow +\infty$:
\begin{eqnarray}
W({\bf F})\sim \frac{1}{1+\alpha}(Gm)^{H}K\frac{1}{F^{H+2}}.
\label{ght}
\end{eqnarray} 
This asymptotic behaviour coincides with the asymptotic behaviour of
the distribution of the force (\ref{nnw6}) due to the nearest
neighbor.

From the asymptotic behaviour (\ref{ght}), the moments $\langle
F^{b}\rangle=\int W({\bf F})F^{b}d{\bf F}$ of the force exist iff
$-2<b<H$. We can obtain an analytical expression for $-2<b<-1/2$. For
convenience, we set $\nu=-b$ with $1/2<\nu<2$. Using Eq. (\ref{de8})
and setting $t=kF$, we have
\begin{eqnarray}
\langle F^{-\nu}\rangle= \int_{0}^{+\infty}dk e^{-ak^{H}}k^{\nu-1} \int_{0}^{+\infty}\frac{J_{0}(t)}{t^{\nu-1}}\, dt. 
\label{de10}
\end{eqnarray}
Using the identity
\begin{eqnarray}
\int_{0}^{+\infty}\frac{J_{0}(t)}{t^{\alpha}}\, dt=\frac{1}{2^{\alpha}}\frac{\Gamma \left (\frac{1-\alpha}{2}\right )}{\Gamma \left (\frac{1+\alpha}{2}\right )},\quad (-\frac{1}{2}<\alpha<1),
\label{de11}
\end{eqnarray}
we obtain for $1/2<\nu<2$:
\begin{eqnarray}
\langle F^{-\nu}\rangle=\frac{1}{2^{\nu-1}Ha^{\nu/H}}\frac{\Gamma\left (\frac{2-\nu}{2}\right )}{\Gamma \left (\frac{\nu}{2}\right )}\Gamma\left (\frac{\nu}{H}\right ).
\label{de12}
\end{eqnarray}

\subsection{The Cauchy distribution}
\label{sec_cauchy}

We note that the characteristic function (\ref{df9}) is linear when
\begin{equation}
H=1,\qquad (\gamma=d+1).
\label{c0}
\end{equation}
In that case, the distribution
of the force is a Cauchy law. This corresponds to $\alpha+p=1$
independently on the dimension of space $d$. For the gravity case
($\alpha=0$), the Cauchy law is obtained for $p=1$, i.e. for a fractal
dimension $d_{f}=d-1$ (assuming $d>1$). In $d$ dimensions, the Cauchy
law has the form
\begin{equation}
W({\bf F})=\frac{a}{\pi^{\frac{d+1}{2}}}\Gamma \left (\frac{1}{2}+\frac{d}{2}\right )\frac{1}{(a^2+F^2)^{\frac{d+1}{2}}}.
\label{c1}
\end{equation}
The moments $\langle F^{b}\rangle$ exist for $-d<b<1$ and they are given by
\begin{equation}
\langle F^{b}\rangle=\frac{a^{b}}{\sqrt{\pi}}\Gamma\left (\frac{1}{2}+\frac{d}{2}\right )\frac{\Gamma\left (\frac{1-b}{2}\right )\Gamma\left (\frac{b+d}{2}\right )}{\Gamma\left (\frac{d}{2}\right )\Gamma\left (\frac{1+d}{2}\right )}.
\label{c1bis}
\end{equation}

For $d=3$, we obtain
\begin{equation}
W({\bf F})=\frac{a}{\pi^{2}}\frac{1}{(a^2+F^2)^{2}},
\label{c2}
\end{equation}
with
\begin{equation}
a=\frac{\pi^{2}}{2+\alpha}GmK.
\label{c3}
\end{equation}
We  have
\begin{equation}
W({\bf F})\rightarrow \frac{1}{\pi^{2}a^3}, \qquad (F\rightarrow 0),
\label{c4}
\end{equation}
\begin{equation}
W({\bf F})\sim \frac{a}{\pi^{2}F^{4}}, \qquad (F\rightarrow +\infty).
\label{c5}
\end{equation}
For $-3<b<1$, the moments are
\begin{equation}
\langle F^{b}\rangle=\frac{a^{b}(b+1)}{\cos\left (\frac{b\pi}{2}\right )}.
\label{c6}
\end{equation}
Using identity (\ref{t16nd}), we check that Eqs. (\ref{c6}),
(\ref{t15}) and (\ref{c1bis}) coincide.

For $d=2$, we obtain
\begin{equation}
W({\bf F})=\frac{a}{2\pi}\frac{1}{(a^2+F^2)^{3/2}},
\label{cd1}
\end{equation}
with
\begin{equation}
a=\frac{2\pi}{\alpha+1}GmK.
\label{cd2}
\end{equation}
We have
\begin{equation}
W({\bf F})\rightarrow \frac{1}{2\pi a^2}, \qquad (F\rightarrow 0),
\label{cd3}
\end{equation}
\begin{equation}
W({\bf F})\sim \frac{a}{2\pi F^{3}}, \qquad (F\rightarrow +\infty).
\label{cd4}
\end{equation}
For $-2<b<1$, the moments are
\begin{equation}
\langle F^{b}\rangle=\frac{a^{b}}{\sqrt{\pi}}\Gamma\left (\frac{1-b}{2}\right )\Gamma\left (\frac{b+2}{2}\right ).
\label{cd5}
\end{equation}

For $d=1$, we obtain
\begin{equation}
W({F})=\frac{a}{\pi}\frac{1}{a^2+F^2},
\label{c6q}
\end{equation}
with
\begin{equation}
a=\frac{\pi}{\alpha}GmK.
\label{c7}
\end{equation}
We have
\begin{equation}
W({F})\rightarrow \frac{1}{\pi a}, \qquad (F\rightarrow 0),
\label{c8}
\end{equation}
\begin{equation}
W({F})\sim \frac{a}{\pi F^{2}}, \qquad (F\rightarrow +\infty).
\label{c9}
\end{equation}
For $-1<b<1$, the moments are
\begin{equation}
\langle F^{b}\rangle=\frac{a^{b}}{\cos(b\frac{\pi}{2})}.
\label{c10}
\end{equation}
Using the identity (\ref{t16nd}), we check that
Eqs. (\ref{c10}), (\ref{c1bis}) and (\ref{u28}) coincide.

\subsection{The marginal Gaussian distribution (critical case)}
\label{sec_margw}

We have seen that the distribution of the total force is a L\'evy law
when condition (\ref{in3}), or equivalently condition (\ref{mon3}), is
fulfilled corresponding to $0<H<2$.  The critical case happens for
$2\alpha+p=2-d$ corresponding to 
\begin{eqnarray}
H=2, \qquad (\gamma=d+2).
\label{hkr}
\end{eqnarray}
In that case, the variance of
the force produced by a star diverges logarithmically. For the gravity
case ($\alpha=0$) this corresponds to $p=2-d$. But, the condition
$d>1$ is required to have a decreasing force and the condition $d\le
2$ must hold to have a non increasing density profile. Therefore, the
only possibility is $d=2$ and $p=0$ treated in Sec. \ref{sec_m}. For
other values of $\alpha$, the dimension of space must lie in the range
$1-\alpha<d\le 2(1-\alpha)$.  When $H=2$, Eq. (\ref{df1}) remains
valid with
\begin{eqnarray}
A({\bf k})=e^{-C({\bf k})}, \qquad C({\bf k})=\int_{|{\bf r}|=0}^{R}
n({\bf r}) \left (1-e^{i {\bf k}\cdot {\bf f}}\right )d{\bf r}.\nonumber\\
\label{wm1}
\end{eqnarray}
Note that the integral defining the characteristic function diverges
logarithmically as $R\rightarrow +\infty$. Therefore, Eq. (\ref{df1})
with Eq. (\ref{wm1}) must be viewed as an equivalent of the
distribution $W_{N}({\bf F})$ for large values of $R$ or $N$, not a
true limit. Equations (\ref{df9})-(\ref{df12}) are now replaced by
\begin{eqnarray}
C({\bf k})=a k^{2}, 
\label{df9b}
\end{eqnarray}
where 
\begin{eqnarray}
a=\frac{S_{d}}{d+\alpha-1}(Gm)^{2}KB,
\label{df11b}
\end{eqnarray}
\begin{eqnarray}
B=\int_{\frac{Gmk}{R^{d+\alpha-1}}}^{+\infty}\frac{dx}{x^{3}}\left \lbrack 1-\Gamma\left (\frac{d}{2}\right )\left (\frac{2}{x}\right )^{\frac{d}{2}-1}J_{\frac{d}{2}-1}(x)\right \rbrack.\nonumber\\
\label{wm2}
\end{eqnarray}
Since this integral diverges logarithmically as $R\rightarrow +\infty$, we can replace the term in brackets by its leading order expression for $x\rightarrow 0$ using Eq. (\ref{df13}) and we obtain
\begin{eqnarray}
B=\frac{1}{2d}\int_{\frac{Gmk}{R^{d+\alpha-1}}}^{+\infty}\frac{dx}{x}=\frac{1}{4d}\ln\left (\frac{N}{G^2 m^2 k^2}\right ),
\label{wm3}
\end{eqnarray}
where we have used $N\sim R^{d-p}\sim R^{2(d+\alpha-1)}$ for
$R,N\rightarrow +\infty$. Redefining 
\begin{eqnarray}
\overline{a}=\frac{S_{d}}{d+\alpha-1}(Gm)^{2}\frac{K}{4d}
\label{wm4}
\end{eqnarray}
the distribution of the force can be written
\begin{eqnarray}
W({\bf F})=\frac{S_{d}}{(2\pi)^{d}}\Gamma\left (\frac{d}{2}\right )\left (\frac{2}{F}\right )^{\frac{d}{2}-1}\nonumber\\
\times\int_{0}^{+\infty} e^{-\overline{a}k^{2}\ln\left (\frac{N}{G^2 m^2 k^2}\right )}k^{\frac{d}{2}}J_{\frac{d}{2}-1}(kF)dk.
\label{wm5}
\end{eqnarray}
For $F$ not too large (corresponding to $k$ not too small), we can replace the logarithmic term by its leading contribution $\ln N$ and we get
\begin{eqnarray}
W({\bf F})=\frac{S_{d}}{(2\pi)^{d}}\Gamma\left (\frac{d}{2}\right )\left (\frac{2}{F}\right )^{\frac{d}{2}-1}\nonumber\\
\times\int_{0}^{+\infty} e^{-\overline{a}k^{2}\ln N}k^{\frac{d}{2}}J_{\frac{d}{2}-1}(kF)dk.
\label{wm6}
\end{eqnarray}
In that case, the distribution of the force is Gaussian 
\begin{eqnarray}
W({\bf F})=\frac{1}{(4\pi \overline{a} \ln N)^{d/2}}e^{-\frac{F^2}{4\overline{a}\ln N}},  
\label{wm7}
\end{eqnarray}
with a variance $\langle F^2\rangle=2d\overline{a}\ln N$ diverging
like $\ln N$. For $F\rightarrow +\infty$ (corresponding to
$k\rightarrow 0$), we can replace the logarithmic term by $-2\ln k$
and we get
\begin{eqnarray}
W({\bf F})=\frac{S_{d}}{(2\pi)^{d}}\Gamma\left (\frac{d}{2}\right )\left (\frac{2}{F}\right )^{\frac{d}{2}-1}\nonumber\\
\times\int_{0}^{+\infty} e^{2\overline{a}k^{2}\ln k}k^{\frac{d}{2}}J_{\frac{d}{2}-1}(kF)dk.
\label{wm8}
\end{eqnarray}
Repeating the procedure of Sec. \ref{sec_df}, we can rewrite the foregoing integral in the form
\begin{eqnarray}
W({\bf F})=\frac{C_{d}}{(2\pi)^{d}F^d}{\rm Re} \int_{C^+}
(1-t^2)^{\frac{d-3}{2}}\, dt\nonumber\\
\times \int_{0}^{+\infty} e^{2\overline{a}\left (\frac{iy}{tF}\right )^2 \ln \left (\frac{iy}{tF}\right )}e^{-y} \left (\frac{i}{t}\right )^{d} y^{d-1}\, dy,
\label{wen1}
\end{eqnarray}
where we recall that the integral on $t$ has to be performed in the
complex plane along the semi-circle of unit radius in the upper half
plane ${\rm Im}(t)>0$. Expanding the exponential term for large $F$, we obtain
\begin{eqnarray}
W({\bf F})=\frac{C_{d}}{(2\pi)^{d}F^d}{\rm Re} \int_{C^+}
(1-t^2)^{\frac{d-3}{2}}\, dt\int_{0}^{+\infty}e^{-y}  \nonumber\\
\times  \left \lbrack 1-2\overline{a}\frac{1}{t^2}\frac{y^2}{F^2} \ln \left (\frac{iy}{F}\right )+2\overline{a}\frac{\ln t}{t^2}\frac{y^2}{F^2}+...\right\rbrack i^{d} \frac{y^{d-1}}{t^d}\, dy.\nonumber\\
\label{wen2}
\end{eqnarray}
Now, setting $t=e^{i\theta}$ and integrating on $C_+$ from $\theta=\pi$ to $\theta=0$, we obtain the following results
\begin{eqnarray}
{\rm Re}\int_{C^+}
(1-t^2)^{\frac{d-3}{2}} \frac{1}{t^{d}}i^d\, dt  =0, 
\label{wen3}
\end{eqnarray}
\begin{eqnarray}
{\rm Re}\int_{C^+}
(1-t^2)^{\frac{d-3}{2}}\frac{1}{t^{d+2}}i^d\, dt =0, 
\label{wen4}
\end{eqnarray}
\begin{eqnarray}
{\rm Re}\int_{C^+}
(1-t^2)^{\frac{d-3}{2}}\frac{1}{t^{d+2}}i^{d+1}\, dt =0, 
\label{wen5}
\end{eqnarray}
\begin{eqnarray}
{\rm Re}\int_{C^+}
(1-t^2)^{\frac{d-3}{2}} \frac{\ln t}{t^{d+2}}i^{d} \, dt\nonumber\\
  =\frac{2^{d-1}d\sqrt{\pi}}{\Gamma(d+2)}\Gamma\left (\frac{d-1}{2}\right )\Gamma\left (\frac{d}{2}\right ). 
\label{wen6}
\end{eqnarray}
Substituting these relations in Eq. (\ref{wen2}) and using Eq. (\ref{wm4}), we  obtain the equivalent for $F\rightarrow +\infty$:
\begin{eqnarray}
W({\bf F})\sim K\frac{(Gm)^{2}}{d+\alpha-1}\frac{1}{F^{d+2}},
\label{wm9}
\end{eqnarray}
which coincides with the distribution (\ref{nnw6}) due to the nearest
neighbor for $H=2$. We also
note that the asymptotic behaviour (\ref{df17}) obtained for $H<2$ passes
to the limit $H\rightarrow 2$. Finally, the crossover between the two distributions (\ref{wm7}) and (\ref{wm9}) occurs for a typical force
\begin{eqnarray}
F_{crit}(N)\sim (4\overline{a}\ln N)^{1/2}\ln (\ln N)^{1/2}.
\label{hekf}
\end{eqnarray} 
in any dimension of space $d$.

It may be useful to study the physical dimensions $d=3,2,1$
specifically and re-derive the previous results in a different
manner. For the physical dimension $d=3$, the expression (\ref{wm5})
of the distribution of the force reduces to
\begin{eqnarray}
W({\bf F})=\frac{1}{2\pi^{2}F}\int_{0}^{+\infty} e^{-\overline{a}k^2\ln\left (\frac{N}{G^2 m^2 k^2}\right )}\sin(kF)k\, dk,\nonumber\\
\label{wm10}
\end{eqnarray}
with
\begin{eqnarray}
\overline{a}=\frac{K\pi (Gm)^{2}}{3(2+\alpha)}.
\label{wm11}
\end{eqnarray}
This critical case corresponds to $2\alpha+p=-1$ or, equivalently, $\gamma=5$.
In the core of the distribution, we can make the approximation
\begin{eqnarray}
W({\bf F})=\frac{1}{2\pi^{2}F}\int_{0}^{+\infty} e^{-\overline{a}\ln N k^2}\sin(kF)k\, dk.
\label{wm12}
\end{eqnarray}
This leads to the Gaussian distribution
\begin{eqnarray}
W({\bf F})=\frac{1}{(4\pi \overline{a}\ln N)^{3/2}}e^{-\frac{F^{2}}{4\overline{a}\ln N}},
\label{wm13}
\end{eqnarray}
with a variance $\langle F^2\rangle=6\overline{a}\ln N$ that diverges
logarithmically with $N$.  In the tail of the distribution, we can make the approximation
\begin{eqnarray}
W({\bf F})=\frac{1}{2\pi^{2}F}\int_{0}^{+\infty} e^{2\overline{a}k^2 \ln k}\sin(kF)k\, dk.
\label{wm14}
\end{eqnarray}
Setting $z=kF$, this can be rewritten
\begin{eqnarray}
W({\bf F})=\frac{1}{2\pi^{2}F^3}{\rm Im}\int_{0}^{+\infty} e^{2\overline{a}\left (\frac{z}{F}\right )^{2}\ln  \left (\frac{z}{F}\right )} e^{iz} z\, dz.
\label{wm15}
\end{eqnarray}
In this expression, $z$ is real and the domain of integration is on
the real axis $z\ge 0$. Under this form, we cannot expand the
exponential in power series for $F\rightarrow +\infty$ and integrate
term by term because the integrals would diverge. However, we can
modify the domain of integration and work in the complex plane. Then,
$z$ is viewed as a complex variable and we can replace the domain of
integration by the imaginary axis, i.e. $z=iy$ with $y$ real $\ge
0$. Thus, we have
\begin{eqnarray}
W({\bf F})=-\frac{1}{2\pi^{2}F^3}{\rm Im}\int_{0}^{+\infty} e^{-2\overline{a}\left (\frac{y}{F}\right )^{2} \ln \left (\frac{iy}{F}\right )} e^{-y} y\, dy.\nonumber\\
\label{wm16}
\end{eqnarray}
Under this form, it is possible to expand the exponential in powers of $y/F$ and integrate term by term. To leading order, we find
\begin{eqnarray}
W({\bf F})=-\frac{1}{2\pi^{2}F^3}{\rm Im}\int_{0}^{+\infty} \biggl\lbrack 1-2\overline{a}\left (\frac{y}{F}\right )^{2}\nonumber\\
\times \ln \left (\frac{iy}{F}\right )+...\biggr \rbrack e^{-y}y\, dy.
\label{wm17}
\end{eqnarray} 
Since only the imaginary part of the integral matters, the foregoing expression
reduces to
\begin{eqnarray}
W({\bf F})\sim \frac{\overline{a}}{\pi^{2}F^5}{\rm Im}\int_{0}^{+\infty} \ln(i) e^{-y} y^3 \, dy.
\label{wm18}
\end{eqnarray} 
Then, using $\ln(i)=\ln(e^{i\pi/2})=i\pi/2$ we finally obtain
\begin{eqnarray}
W({\bf F})\sim \frac{\overline{a}}{2\pi F^5}\int_{0}^{+\infty} e^{-y} y^3 \, dy,
\label{wm19}
\end{eqnarray} 
which yields
\begin{eqnarray}
W({\bf F})\sim \frac{\overline{a}}{2\pi F^5}\Gamma(4)\sim  \frac{3\overline{a}}{\pi F^5}.
\label{wm20}
\end{eqnarray} 
We obtain the same result if we deform the contour of integration in
Eq. (\ref{wm15}) as indicated in Chandrasekhar \cite{c48} for the case
$H<2$.  Substituting the value of $\overline{a}$ from
Eq. (\ref{wm11}), we obtain the equivalent
\begin{eqnarray}
W({\bf F})\sim \frac{K(Gm)^{2}}{2+\alpha}\frac{1}{F^{5}}, \qquad (F\rightarrow +\infty) 
\label{wm21}
\end{eqnarray} 
which is in exact agreement with the distribution of the force
(\ref{nnw6}) due to the nearest neighbor for $H=2$ and $d=3$. We also
note that the asymptotic behaviour (\ref{t14}) obtained for $H<2$ passes
to the limit $H\rightarrow 2$. Finally, the crossover between the two distributions (\ref{wm13}) and (\ref{wm21}) occurs for a typical force
\begin{eqnarray}
F_{crit}(N)\sim (4\overline{a}\ln N)^{1/2}\ln (\ln N)^{1/2}.
\label{wm22}
\end{eqnarray}

For the dimension $d=1$, the expression (\ref{wm5}) of the
distribution of the force reduces to
\begin{eqnarray}
W({F})=\frac{1}{\pi}\int_{0}^{+\infty} e^{-\overline{a}k^2\ln\left (\frac{N}{G^2 m^2 k^2}\right )}\cos(kF)\, dk,\nonumber\\
\label{wm23}
\end{eqnarray}
with
\begin{eqnarray}
\overline{a}=\frac{K(Gm)^{2}}{2\alpha}.
\label{wm24}
\end{eqnarray}
This critical case corresponds to $2\alpha+p=1$ or, equivalently,
$\gamma=3$. In the core of the distribution, we can make the
approximation
\begin{eqnarray}
W({F})=\frac{1}{\pi}\int_{0}^{+\infty} e^{-\overline{a}\ln N k^2}\cos(kF)\, dk.
\label{wm25}
\end{eqnarray}
This leads to the Gaussian distribution
\begin{eqnarray}
W({F})=\frac{1}{(4\pi \overline{a}\ln N)^{1/2}}e^{-\frac{F^{2}}{4\overline{a}\ln N}},
\label{wm26}
\end{eqnarray}
with a variance $\langle F^2\rangle=2\overline{a}\ln N$ that diverges
logarithmically with $N$.  In the tail of the distribution, we can make the approximation
\begin{eqnarray}
W({F})=\frac{1}{\pi}\int_{0}^{+\infty} e^{2\overline{a}k^2 \ln k}\cos(kF)\, dk.
\label{wm27}
\end{eqnarray}
Setting $z=kF$, this can be rewritten
\begin{eqnarray}
W({F})=\frac{1}{\pi F}{\rm Re}\int_{0}^{+\infty} e^{2\overline{a}\left (\frac{z}{F}\right )^{2}\ln  \left (\frac{z}{F}\right )} e^{iz} \, dz.
\label{wm28}
\end{eqnarray}
To determine the asymptotic behaviour of $W(F)$ for $F\rightarrow
+\infty$, we deform the contour of integration and integrate on the
imaginary axis, setting $z=iy$ with $y$ real $\ge 0$. Thus, we have
\begin{eqnarray}
W({F})=\frac{1}{\pi F}{\rm Re}\int_{0}^{+\infty} i e^{-2\overline{a}\left (\frac{y}{F}\right )^{2} \ln \left (\frac{iy}{F}\right )} e^{-y} \, dy.\nonumber\\
\label{wm29}
\end{eqnarray}
We can now expand the exponential in powers of $y/F$ and integrate
term by term. To leading order, we find
\begin{eqnarray}
W({F})=\frac{1}{\pi F}{\rm Re}\int_{0}^{+\infty} i \biggl\lbrack 1-2\overline{a}\left (\frac{y}{F}\right )^{2}\nonumber\\
\times \ln \left (\frac{iy}{F}\right )+...\biggr \rbrack e^{-y}\, dy.
\label{wm30}
\end{eqnarray} 
Since only the real part of the integral matters, the foregoing expression
reduces to
\begin{eqnarray}
W({F})\sim -\frac{2\overline{a}}{\pi F^3}{\rm Re}\int_{0}^{+\infty} i \ln(i) e^{-y} y^2 \, dy.
\label{wm31}
\end{eqnarray} 
Then, using $\ln(i)=\ln(e^{i\pi/2})=i\pi/2$ we finally obtain
\begin{eqnarray}
W({F})\sim \frac{\overline{a}}{F^3}\int_{0}^{+\infty} e^{-y} y^2 \, dy,
\label{wm32}
\end{eqnarray} 
which yields
\begin{eqnarray}
W({F})\sim \frac{\overline{a}}{F^3}\Gamma(3)\sim  \frac{2\overline{a}}{F^3}.
\label{wm33}
\end{eqnarray} 
We obtain the same result if we deform the contour of integration in
Eq. (\ref{wm28}) as indicated between Eqs. (\ref{u12}) and (\ref{u18})
for the case $H<2$.  Substituting the value of $\overline{a}$ from
Eq. (\ref{wm24}), we obtain the equivalent
\begin{eqnarray}
W({F})\sim \frac{K(Gm)^{2}}{\alpha}\frac{1}{F^{3}}, \qquad (F\rightarrow +\infty) 
\label{wm34}
\end{eqnarray} 
which is in exact agreement with the distribution of the force
(\ref{nnw6}) due to the nearest neighbor for $H=2$ and $d=1$. We also
note that the asymptotic behaviour (\ref{u19}) obtained for $H<2$ passes
to the limit $H\rightarrow 2$. Finally, the crossover between the two distributions (\ref{wm26}) and (\ref{wm33}) occurs for a typical force
\begin{eqnarray}
F_{crit}(N)\sim (4\overline{a}\ln N)^{1/2}\ln (\ln N)^{1/2}.
\label{wm35}
\end{eqnarray} 
The critical case in $d=2$ corresponding to $2\alpha+p=0$ or, equivalently $\gamma=4$, can be treated like in \cite{cs1}.

\section{Conclusion}
\label{sec_conc}

In this paper, we have studied how the statistics of the gravitational
force created by a random distribution of field sources changes with
the dimension of space $d$. The dimensions $d=1$, $d=2$ and $d=3$
correspond respectively to plane-parallel (sheets), cylindrical
(filaments) and spherical (stars) configurations. We have shown that
the dimension $d=2$ is critical\footnote{The critical nature of the
dimension $d=2$ has also been noted in \cite{sc} regarding the
gravitational collapse of isothermal systems.} as it separates
Gaussian laws (for $d=1$) from L\'evy laws (for $d\ge 3$). This
transition may have interesting implications for the kinetic theory of
self-gravitating systems (in $d$ dimensions) since the distribution of
the gravitational force is a key ingredient for the determination of
the diffusion coefficient of stars \cite{kandrup,kind}. Furthermore, even
if our study has astrophysical motivations at the start, it can be of
interest in probability theory to illustrate the differences between
Gaussian and L\'evy laws.

Note, however, that our analysis is based on several simplifying
assumptions: 

(i) we have assumed (for the cases $d=3$ and $d=2$) that the number of
stars $N\rightarrow +\infty$ or equivalently that the system is
infinite, i.e. we have considered the limit $N,R\rightarrow +\infty$
with fixed $n=N/V$. The distribution of the force $W_N({\bf F})$ for a
finite system must be computed numerically, using
Eqs. (\ref{h5})-(\ref{h6}). In $d=3$ this study has been performed by
Ahmad \& Cohen \cite{ac}. It is shown that the convergence to the
limit distribution $W({\bf F})$ is quite rapid: for $N=2$, the overall
agreement with the $N=\infty$ case is not unreasonable, for $N=50$ the
agreement is accurate to $10\%$ and for $N=1000$, the agreement is
excellent. In $d=2$, the convergence of the distributution of the
gravitational force to the limit distribution (Gaussian) is very slow
and, as we have seen, a power-law tail develops at a typical value of the force
(\ref{m12}) which increases logarithmically with $N$. Numerical
simulations exhibiting this power-law tail are reported (for point
vortices) in
\cite{jimenez,min,kuv,lm}. In $d=1$, the exact distribution of the
gravitational force has been obtained for any $N$.

(ii) we have assumed in Secs. \ref{sec_h}-\ref{sec_g} that the distribution of
stars is spatially homogeneous. In fact, an infinite and homogeneous
distribution of stars is not stable and it clusters in dense objects
(galaxies or clusters of galaxies). The case of a power-law
distribution of stars has been treated by Kandrup \cite{kandrup} in
$d=3$ and generalized to any dimension and to any power-law force in
Sec. \ref{sec_in}. In practice, since the distribution of the force is
dominated by the nearest neighbor, the assumption of an infinite and
homogeneous distribution is not crucial; indeed, Kandrup \cite{ka}
shows that the inhomogeneous case, for a smooth density distribution
$n({\bf r})$, is still described by the Holtsmark distribution (for
the fluctuating force) where the density $n$ is replaced by the local
density $n({\bf a})$ at the point under consideration (see also
Appendix A of
\cite{kinvortex} in $d=2$).

(iii) we have assumed that the positions of the stars are uncorrelated
(Poisson distribution). This approximation may be correct in stellar
dynamics where it is known that the two-body distribution function can
be approximated by a product of two one-body distributions\footnote{Of course, for finite $N$ systems, correlations must be
taken into account in the so-called ``collisional'' regime of the
dynamics, as they drive the kinetic evolution of the system
\cite{saslaw,kandrupkin,chavkin}.} in a proper thermodynamic limit $N\rightarrow +\infty$
\cite{metastable}. However, in that case, the one-body distribution is
spatially inhomogeneous and we are led to point (ii). On the other
hand, in cosmology, the system is statistically spatially homogeneous
but the particles are correlated and have the tendency to form
clusters. In that case, the theoretical framework developed to
determine the statistics of the gravitational force must be
modified. Some interesting attempts to take into account spatial
correlations in the position of the particles have been made in
\cite{ad,gab}. Therefore, the results presented in this paper 
can be improved in several directions by relaxing the above
assumptions. This will be considered in future works.

A last comment, suggested by the referee, may be in order. The 
limit distribution for a sum of random variables ${\bf
F}=\sum_{i=1}^{N}{\bf f}_i$ is a classical problem in probability
theory and there exists rigorous results and general theorems about it
\cite{levy,gk,feller,bg,bardou,sornette}. 
Our approach, which is based on the seminal work of Chandrasekhar
\cite{chandra}, is consistent with these general theorems. 
In fact, Chandrasekhar (1943)
\cite{chandra} and Holtsmark (1919) \cite{holtsmark} obtained ``L\'evy laws''
independently from L\'evy (1937) \cite{levy} and other mathematicians
of that time. No reference to L\'evy laws are made in the classical
papers of Chandrasekhar (1943, 1948) \cite{chandra,c48}, nor in the
more recent review of Kandrup (1980) \cite{kandrup}. Reciprocally, the
books
\cite{gk,feller,bg,bardou} do not make any reference to
Chandrasekhar's work and derive the limit distributions in a more
formal manner. In the present paper, we have used and extended the
method introduced by Chandrasekhar \cite{chandra}. One interest of
this method is that it is fully explicit and amounts to the
calculation of integrals. It is, however, restricted to {\it pure}
power-law distributions $\tau({\bf f})\propto f^{-\gamma}$ (with a
cut-off at small $f$) while the theorems of \cite{gk,feller,bg,bardou}
are more general. We have obtained $d$-dimensional generalizations of
L\'evy laws [see Eq. (\ref{df16})] and given their main properties
(usually, the problem of the sum of random variables is formulated in
one dimension).  On the other hand, the critical case
$\gamma=\gamma_{c}=d+2$ reported in Secs. \ref{sec_m} and
\ref{sec_margw} of the present paper (where the variance of the random
variables $\langle f^2\rangle$ diverges logarithmically) has not been
treated in depth in
\cite{gk,feller,bg,bardou}. It is usually argued that, in that case,
the limit distribution $W({\bf F})$ is Gaussian.  This is true in a
strict sense when $N\rightarrow +\infty$. However, we have shown that
for large but finite $N$, the physical distribution $W({\bf F})$ has a
Gaussian core and an algebraic tail. The separation between these two
behaviours is obtained for a typical value of the force $F_c(N)\propto
(\ln N)^{1/2}\ln(\ln N)^{1/2}$ which diverges with $N$. Therefore, at
the limit $N\rightarrow +\infty$, the power-law tail is rejected to
infinity and only the Gaussian core remains. However, the convergence
is so slow (logarithmic) that, in practice, the power-law tail is
visible. This point may have been overlooked in
\cite{gk,feller,bg,bardou}.

Finally, our study of the statistics of the gravitational force
created by a uniform distribution of sources in $d$ dimensions
illustrates the three kinds of laws that can be read from Fig. 1.1. of
the review of Bouchaud \& Georges \cite{bg}: (i) For $d>d_c=2$, the
variance of the individual forces diverges algebraically since
$\gamma=d^2/(d-1)<\gamma_{c}=2+d$ [$\gamma=9/2<\gamma_c=5$ in $d=3$]
and the distribution of the total force is a $d$-dimensional L\'evy
law (\ref{df16}) with index $H=d/(d-1)$ [$H=3/2$ in $d=3$]. (ii) For
$d<d_c=2$, the variance of the individual forces is finite and the
distribution of the total force is Gaussian according to the CLT (in
$d=1$ it is exactly given by a Bernouilli law for all $N$). (iii) For
$d=d_c=2$ (critical case), the variance of the individual forces
diverges logarithmically since $\gamma=\gamma_c=4$ and the
distribution of the total force is a marginal Gaussian
distribution. In that case, we are at the border between Gaussian and
L\'evy laws (see again Fig. 1.1. of
\cite{bg}). More generally, our results can be expressed in 
terms of $\gamma$ alone (in a space of dimension $d$). For
$d<\gamma<\gamma_{c}=d+2$, the variance of the individual forces
diverges algebraically and the distribution of the total force is a
$d$-dimensional L\'evy law (\ref{df16}) with index $H=\gamma-d$ (its
tail decreases algebraically with an exponent $\gamma$ due to the nearest
neighbour). For $\gamma>\gamma_{c}$, the variance of the individual
forces is finite and the distribution of the total force is Gaussian
according to the CLT.  For $\gamma=\gamma_c$ (critical case), the
variance of the individual forces diverges logarithmically and the
distribution of the total force is a marginal Gaussian distribution. 
Therefore, for a fixed dimension of space $d$, the different laws correspond
to $d<\gamma<\gamma_{c}$ (L\'evy), $\gamma=\gamma_{c}$ (marginal) and $\gamma>\gamma_{c}$ (Gaussian). Alternatively, when we consider the gravitational force created by a homogeneous distribution of sources, $\gamma=d^2/(d-1)$ is fixed and the different laws  correspond to $d>d_c=2$
(L\'evy), $d=d_c$ (critical) and $d<2$ (Gaussian).

\appendix

\section{The distribution of the gravitational force created by the nearest neighbor in $d$ dimensions}
\label{sec_pvg}

In this Appendix, we determine the distribution of the gravitational
force in $O$ due to the contribution of the nearest neighbor for a
uniform distribution of stars. For the sake of generality, we work in a
space of $d$ dimensions. The probability $\tau_{n.n.}(r)dr$ that the
position of the nearest neighbor occurs between $r$ and $r+dr$ is
equal to the probability that no star exist interior to $r$ times the
probability that a star (any) exists in the shell between $r$ and
$r+dr$. Therefore, it satisfies an equation of the form
\begin{equation}
\tau_{n.n.}(r)dr=\left (1-\int_{0}^{r}\tau_{n.n.}(r')dr'\right )nS_{d}r^{d-1}dr,
\label{pvg1}
\end{equation}
where $n$ is the mean density of stars. Differentiating this
expression with respect to $r$, we obtain
\begin{equation}
\frac{d}{dr}\left\lbrack \frac{\tau_{n.n.}(r)}{S_{d}n r^{d-1}}\right\rbrack=-\tau_{n.n.}(r).
\label{pvg2}
\end{equation}
This equation is readily integrated with the condition $\tau_{n.n.}(r)\sim S_{d}n r^{d-1}$ for $r\rightarrow 0$ and we find
\begin{equation}
\tau_{n.n.}(r)=S_{d}n r^{d-1}e^{-\frac{S_{d}nr^{d}}{d}}.
\label{pvg3}
\end{equation}
From this formula, we can obtain the exact expression of the ``average distance'' $D$  between stars:
\begin{eqnarray}
D=\int_{0}^{+\infty}\tau_{n.n.}(r)r\, dr
=\int_{0}^{+\infty}S_{d}n r^{d}e^{-\frac{S_{d}nr^{d}}{d}}\, dr\nonumber\\
=\left (\frac{d}{S_{d}n}\right )^{1/d}\int_{0}^{+\infty}x^{1/d}e^{-x}dx
=\left (\frac{d}{S_{d}n}\right )^{1/d}\Gamma\left (1+\frac{1}{d}\right ).\nonumber\\
\label{pvg4}
\end{eqnarray}
For example, $D=\left (\frac{3}{4\pi n}\right )^{1/3}\Gamma(4/3)$ in $d=3$, $D=\frac{1}{2\sqrt{n}}$ in $d=2$ and $D=\frac{1}{2n}$ in $d=1$.

The gravitational force created in $O$ by a star in ${\bf r}$ in a $d$-dimensional space is given by
\begin{equation}
{\bf F}=Gm\frac{{\bf r}}{r^d}.
\label{pvg5}
\end{equation}
We note that the variance of the force created by one star
\begin{equation}
\langle F^2\rangle\propto \int_{0}^{+\infty} \frac{1}{r^{2(d-1)}}r^{d-1}dr\propto  \int_{0}^{+\infty} \frac{1}{r^{d-1}}dr
\label{pvg6}
\end{equation}
diverges for $d\ge 2$ due to the behaviour at small distances
$r\rightarrow 0$. In that case, the distribution of the (total)
gravitational force is a L\'evy law. For large field strengths, it is
well-approximated by the distribution created by the nearest neighbor. The
distribution of the force created by the nearest neighbor is such that
$W_{n.n.}({\bf F})d{\bf F}=\tau_{n.n.}({\bf r})d{\bf r}$ where
\begin{equation}
\tau_{n.n.}({\bf r})=\frac{\tau_{n.n.}({r})}{S_{d}r^{d-1}}=ne^{-\frac{S_{d}nr^{d}}{d}}.
\label{pvg7}
\end{equation}
Using
\begin{equation}
d{\bf F}=(d-1)(Gm)^{-d/(d-1)}F^{d^2/(d-1)}d{\bf r},
\label{pvg8}
\end{equation}
we obtain
\begin{equation}
W_{n.n.}({\bf F})=n\frac{(Gm)^{d/(d-1)}}{(d-1)F^{d^2/(d-1)}}e^{-\frac{S_{d}n}{d}\left (\frac{Gm}{F}\right )^{d/(d-1)}}.
\label{pvg9}
\end{equation}
For $F\rightarrow +\infty$, we get
\begin{equation}
W_{n.n.}({\bf F})\sim  n\frac{(Gm)^{d/(d-1)}}{(d-1)F^{d^2/(d-1)}}.
\label{pvg10}
\end{equation}
The distribution of the force decreases with an exponent
\begin{equation}
\gamma=d+\frac{d}{d-1}.
\label{pvg10b}
\end{equation}
We note that the average force $\langle F\rangle$ is always finite  and its value is
\begin{equation}
\langle F\rangle_{n.n.}=\left (\frac{S_d}{d}\right )^{\frac{d-1}{d}}\Gamma\left (\frac{1}{d}\right )Gmn^{\frac{d-1}{d}}.
\label{pvg11}
\end{equation}
It can be expressed in terms of the exact average distance between
stars (\ref{pvg4}) under the form
\begin{equation}
\langle F\rangle_{n.n.}=\frac{1}{d^{d-1}}\Gamma\left (\frac{1}{d}\right )^d \frac{Gm}{D^{d-1}}.
\label{pvg12}
\end{equation}


\end{document}